\documentclass[manuscript,screen]{acmart}

\usepackage{subfig}
\usepackage{multirow}

\AtBeginDocument{%
  \providecommand\BibTeX{{%
    \normalfont B\kern-0.5em{\scshape i\kern-0.25em b}\kern-0.8em\TeX}}}

\renewcommand\footnotetextcopyrightpermission[1]{}

\setcopyright{none}

\newcommand{\etal}{\emph{et~al.}}
\newcommand{\ie}{i.e.,}
\newcommand{\eg}{e.g.,}

\newcommand{\mysubsubsection}[1]{\noindent\textbf{#1}}

\begin{document}

\title[Structural Properties and Resilience of Blockchain Overlays]
{A First Look into the Structural Properties and Resilience of Blockchain Overlays}

\author{Aristodemos Paphitis}
\affiliation{%
 \institution{Cyprus University of Technology}
 \city{Limassol}
 \country{Cyprus}}
 \email{am.paphitis@edu.cut.ac.cy}

\author{Nicolas Kourtellis}
\affiliation{%
 \institution{Telefonica Research}
 \city{Barcelona}
 \country{Spain}}
 \email{nicolas.kourtellis@telefonica.com}

\author{Michael Sirivianos}
\affiliation{%
 \institution{Cyprus University of Technology}
 \city{Limassol}
 \country{Cyprus}}
\email{michael.sirivianos@cut.ac.cy}

\renewcommand{\shortauthors}{Paphitis et al.}

\hyphenation{block-chain}
\hyphenation{block-chains}

\begin{abstract}
Blockchain (BC) systems are highly distributed peer-to-peer networks that offer an alternative to centralized services and promise robustness to coordinated attacks.
However, the resilience and overall security of a BC system rests heavily on the structural properties of its underlying peer-to-peer overlay.
Despite their success, BC overlay networks' critical design aspects, connectivity properties and network-layer inter-dependencies are still poorly understood. 
In this work, we set out to fill this gap by studying the most important overlay structural properties and the resilience to targeted attacks of seven distinct BC networks.

In particular, we probe and crawl these BC networks 
over 28 days. 
We construct, at frequent intervals, connectivity graphs for each BC network, consisting of all potential connections between peers. 
We analyze the structural graph properties of these networks, their temporal characteristics, and their topological resilience.
We show that by targeting fewer than 10 highly-connected peers, major BCs such as Bitcoin can be partitioned into disjoint, i.e., disconnected, components.
Finally, we uncover a hidden overlap 
between different BC networks, where certain peers participate in more than one BC network.
This finding has serious implications for the robustness of the overall BC network ecosystem.
\end{abstract}
\maketitle
\section{Introduction}
The widespread adoption of Bitcoin set the stage for the emergence of numerous blockchains (BC). The unique features of the BC technology have increased its visibility and are expected to bring disruptive innovation to many sectors that traditionally rely on centralized, trusted third-parties. This, in turn, raises the question of whether their transport layer infrastructure ensures sufficient resilience and performance.

BCs use structured or unstructured peer-to-peer (P2P) networks that employ stateful connections.
Such P2P overlays are easily constructed, enable fast diffusion of information, and exhibit highly dynamic network topologies. BC-based applications are highly depended on these overlay networks.
The overlay network's properties define the level of BC security, scalability, and resilience.
It is therefore important to analyze these networks to unveil possible limitations and vulnerabilities.
Unfortunately, BC networks are not sufficiently documented by their development teams.
Furthermore, not much research effort has been devoted to study the structural properties of these networks.

The main reason BC networks are understudied is that accurate topology inference is hard and remains an unsolved problem (see Section~\ref{sec:related}). Previously suggested methods are either not applicable anymore due to changes in the reference clients~\cite{coinscope,DBLP:conf/ccs/BiryukovKP14,bitcoin_pr7125}, or infeasible due to transaction fees~\cite{btctiminganalysis,GrundmannNH18,txprobe}. 
At the same time, these methods are impractical to run on multiple BC networks since they require maintaining connections to a high number of peers. Furthermore, Bitcoin core developers constantly update protocol subtleties to prevent leakage of information that would ease topology inference. 

Our key insight is to circumvent the topology inference problem by seeking to uncover possible connections between pairs of nodes (peers). A connection between two nodes is considered possible if one node includes the other in its list of known addresses. By doing this we trade accuracy for completeness;
we synthesize connectivity graphs that include the vast majority of all potential links between the nodes we observe, including the actual connections (see Section~\ref{sec:methodology}). Since the goal of our study is to uncover structural deficiencies in the overlays, our rationale is as follows: the actually realized topology of an overlay is highly unlikely to be resilient if our inferred topology of possible connections is not.

We proceed to study the structural properties that characterize these graphs and their temporal dynamics (see Sections~\ref{sec:results-net} and \ref{sec:results-temp}). We identify several critical BC nodes, whose removal can cause major topological disruption (see Section~\ref{sec:results-robust}). 
Overall, our work investigates the following research questions:
\begin{itemize}
    \item[\textbf{RQ1:}]\label{rq:1} What are the structural and network characteristics of BC overlay networks? 
    \item[\textbf{RQ2:}]\label{rq:2} What is the relationship between different BCs, due to peers and links concurrently participating in multiple BCs? 
    \item[\textbf{RQ3:}]\label{rq:4} How do BC networks evolve through time? 
    \item[\textbf{RQ4:}]\label{rq:3} What are the implications of the network' properties with respect to resilience against targeted attacks? 
\end{itemize}

With this work, we make the following contributions:
\begin{itemize}
    \item We propose an effective and efficient method for studying the topological characteristics of BC networks.
    \item We analyze the network characteristics of seven distinct BC overlay networks over a period of 4 weeks.
    \item Our results show that these BC networks vary in the way they are structured and are highly dynamic through time.
    They belong to the general exponential family of graphs, but are not substantially related to well-known networks like the Internet topology, the Web or social networks, nor do they resemble random networks.
    \item We find a significant number of nodes participating in more than one BC, at the same time (\ie~\textit{overlapping nodes}).
    Their presence is consistent through time and across major BCs.
    They constitute 10-40\% of any BC network, revealing vulnerabilities in the BC ecosystem, as attackers could focus their effort on these nodes to simultaneously disrupt multiple BC networks.
    \item By performing a longitudinal analysis, we discover that nodes with high up-times are also those with the highest degree. This could be exploited by attackers to identify important nodes in networks that base their resilience on topology hiding techniques.
    \item We also investigate the topology robustness of each BC network. We find that by just removing the top 5 central nodes, we get a significant shrinking of the largest connected component (LCC) in major BCs, suggesting that network partitions can be easily performed by a motivated adversary.
    We also observe a noticeable increase of the network diameter and dramatic decrease in the LCC's size by removing less than 10\% of peers.
    Thus, a powerful DDoS attack targeting a few hundred nodes can lead to the collapse of major BC networks.
\end{itemize}
\section{Background} \label{sec:studied-bcs}
In this section we provide background information on the blockchain networks under study. We chose seven BC networks. All are consistently included in the top 50 cryptocurrencies by market capitalization, according to~\cite{coinmarketcap} for the past year.
We list them alphabetically:
\begin{enumerate}
\item \textbf{Bitcoin}~\cite{nakamoto} was the first cryptographic currency to gain widespread adoption. 
\item \textbf{Bitcoin Cash}~\cite{bch, bchjs} is a hard fork of Bitcoin with an increased block size, aiming at increasing transaction throughput and reducing clearance delays in comparison to Bitcoin.
\item \textbf{Dash}~\cite{dash} is another fork of Bitcoin.
It employs a two-tiered network, consisting of mining nodes (peers) and master nodes. This architecture enables very fast transaction confirmation times.
\item \textbf{Dogecoin}~\cite{markus2013dogecoin, young2018dogecoin} is a fork of Litecoin (see below), that yields faster (only one minute) block generation times.
Although it initially started as a joke to satirize the hype surrounding cryptocurrencies, it has gained visibility and high market capitalization~\cite{dogenyt}.
\item \textbf{Ethereum}~\cite{ethereum} is tailored to executing smart contracts for decentralized applications. It is the most well known cryptocurrency after Bitcoin and has the second highest market capitalization.
The main difference with Bitcoin, and its most prominent feature, is the use of a Turing-complete programming language that allows the creation of smart contracts.
\item \textbf{Litecoin}~\cite{ltc} is one of Bitcoin's first forks.
Its differentiating functions include a decreased block generation time of 2.5 minutes and use of a distinct hashing algorithm, Scrypt~\cite{scrypt}. 
\item \textbf{Zcash}~\cite{zec} is a cryptocurrency focused on user privacy and anonymity based on zero knowledge proofs for transaction processing.
\end{enumerate}
With the exception of Ethereum, the aforementioned BCs are descendants of Bitcoin using very similar overlay implementations. Next, we explain the fundamentals of the overlay network of Bitcoin and Ethereum.

\vspace*{-3mm}
\subsection{Bitcoin Overlay Network} \label{sec:btcp2p}
In the Bitcoin overlay network, nodes communicate through non-TLS TCP connections to form an unstructured P2P network. 
Bitcoin's security heavily depends on the global consistent state of the BC, which relies on its Proof-of-Work based consensus protocol.
The communication protocol is briefly documented in~\cite{btcp2p}, but a formal specification does not exist. To understand its subtleties we looked into previous studies \cite{Biryukov19, Neudecker2019, GrundmannNH18} and the Bitcoin Core reference client source code~\cite{btcsourcecode}. 

When a node joins the network for the first time, it queries a set of DNS seeds that are hardcoded in the reference client (Bitcoin Core). The response to this lookup query includes one or more  
IP addresses of full nodes that can accept new incoming connections. Once connected to the network, a node receives unsolicited \texttt{addr} messages from its connected peers, that contain IP addresses and port numbers of other peers in the network. Additionally, the client can send to peers \texttt{getaddr} messages to gather additional peers. The transmitting node can use those IP addresses to quickly update its database of available nodes rather than waiting for new unsolicited \texttt{addr} messages. The reply to a \texttt{getaddr} message may contain up to a 1000 peer addresses. All known addresses are maintained in an in-memory data structure managed by the address manager(\texttt{ADDRMAN}), and are periodically dumped to disk, in the \texttt{peers.dat} file.
This allows the client to connect directly to those peers on subsequent startups without having to use DNS seeds.
Each node maintains in its data structures up to 81920 peer IP addresses.

\subsection{Ethereum Overlay network} \label{sec:ethp2p}
\mysubsubsection{Peer protocols.} Ethereum’s network communication comprises three distinct protocols. \textit{RLPx} serves as a transport protocol and is used for node discovery and establishment of secure communication. \textit{DEVP2P} is used to create the application session. Last, the Ethereum application-level protocol (eth sub-protocol)
facilitates the exchange of BC information between peers, like transactions and blocks.
DEVP2P is not meant to be specific to Ethereum; other BC or network applications can be built on it. 
It supports discovery of other participants and secure communication among them, on top of the RLPx transport protocol. All protocols are documented in Ethereum's official documentation~\cite{ethereum-devp2p}.

\mysubsubsection{Node Discovery.}
Ethereum's node discovery takes place over UDP, while the rest of communication is done through TCP TLS channels.
\textit{RLPx} implements node discovery based on the routing algorithm of Kademlia, a distributed hash table (DHT)~\cite{kademlia}. 
In Ethereum, each peer has a unique 512-bit node ID. A bitwise XOR is used to compute a distance between two Node IDs. Nodes maintain 256 buckets, each containing a number of entries. Each node assigns known peers to a bucket, according to the XOR distance from itself. In order for a new node to find peers, it first adds a hard-coded set of bootstrap node IDs to its routing table. It then sends to these bootstraping nodes a \texttt{FIND\_NODE} message that specifies a random target node ID. Each peer responds with a list of 16 nodes from its own routing table that are closest to the requested target. Subsequently, the node tries to establish a number of connections (typically between 25 and 50) to other peers in the network.
The Ethereum client implementations impose a 4 second delay between successive \texttt{FIND\_NODE} messages from the same node.
\vspace*{-2mm}
\section{Related Work}
\label{sec:related}
Biryukov \etal~\cite{DBLP:conf/ccs/BiryukovKP14} suggested a method for topology discovery by sending fake marker IP addresses to remote peers. Miller \etal~\cite{coinscope} were the first to successfully infer Bitcoin's public network topology. They discovered links between nodes by leveraging timestamps included in \texttt{ADDR} messages.
In their work, as in ours, they found indications that the Bitcoin network is not purely random. 

Delgado-Segura \etal~\cite{txprobe} inferred Bitcoin's network topology using orphaned transactions.  Their method relies on subtleties of Bitcoin's transaction propagation behavior. It involves fabricating double--spending transaction pairs and sending them to a part of the network. Since their method could interfere with ordinary transactions, they have only performed measurements in Bitcoin's testnet. Their results also indicate that Bitcoin's testnet does not resemble a random graph. Using this method with current Bitcoin prices and 11,000 reachable Bitcoin nodes would require more than \$5,700 in transaction fees and more than 12 hours to cover the whole network~\cite{bitcoinfees}.

Neudecker \etal~\cite{btctiminganalysis} used timing analysis of transaction propagation delays, as observed by a monitoring node, to infer topology. Their approach requires a highly connected monitoring node and creation of transactions. Grundmann \etal~\cite{GrundmannNH18}, proposed mechanisms for Bitcoin topology inference based on double--spending transactions. 
However, this method was not intended to perform full network topology inference due to the high cost; with current transaction fees, inferring the connections of a single peer would cost around \$78~\cite{bitcoinfees}.
By exploiting block relay mechanisms, Daniel \etal~\cite{mapz} presented a passive method to infer connections of mining nodes and their direct neighbors in the ZCash network. Neudecker and Hartenstein~\cite{NeudeckerH19} surveyed the network layer of permissionless BCs, simulated a passive method to infer the network topology with substantial accuracy, and highlighted that network topology hiding is an intermediate security requirement.
Finally, work from Dotan \etal~\cite{pingoletSOK} presents a structured overview of BC P2P overlay networks.
Their work highlights differences and commonalities with traditional networks and identifies open research challenges in network design for decentralized systems.

To hinder attacks that utilize topology inference, Bitcoin Core developers implemented a series of changes in the network protocol. To mitigate the methods described in ~\cite{DBLP:conf/ccs/BiryukovKP14}, the Bitcoin client now drops \texttt{GETADDR} requests from inbound connections ~\cite{againstBiryukov}. To address the adversarial methods proposed by Miller et al.~\cite{coinscope}, nodes stopped updating the timestamp field in the address manager, making it impossible to infer active connections~\cite{againstMillerandBKB}. Neudecker's timing analysis is also rendered impractical due to code changes~\cite{bitcoin_pr7125}. 

Despite previous efforts, very little is known regarding the structure and topological properties of BC overlay networks. Instead, past studies focused on methods for inferring the well--hidden topology of Bitcoin, either against the whole network or a specific peer. With the exception of~\cite{coinscope}, these studies were validated against the Bitcoin testnet~\cite{txprobe}, or against selected nodes~\cite{btctiminganalysis, GrundmannNH18}. We also note that methods described in \cite{txprobe, GrundmannNH18, DBLP:conf/ccs/BiryukovKP14} have ethical issues since they actively send fake information or malformed transactions to the network.

To the best of our knowledge, this is the first study that focuses on the network structural properties of multiple BC networks. As we show in Section~\ref{sec:methodology-assessment}, we circumvent the challenges of topology inference and build a simple network monitor that can probe seven different BC networks in parallel to uncover the vast majority of potential connections. Our implementation does not require high connectivity in each network and is free of transaction processing costs, allowing for greater scalability. Similar to~\cite{coinscope}, our work does not interfere with transactions and only uses information made available by the network peers.
Consequently, we are the first to study the crawled BC networks in--depth for their network characteristics and properties. Furthermore, we investigate the resilience of these networks against random and targeted attacks.
\section{Methodology} \label{sec:methodology} 

\begin{figure}[t]
\centering
  {\includegraphics[width=\linewidth, clip]{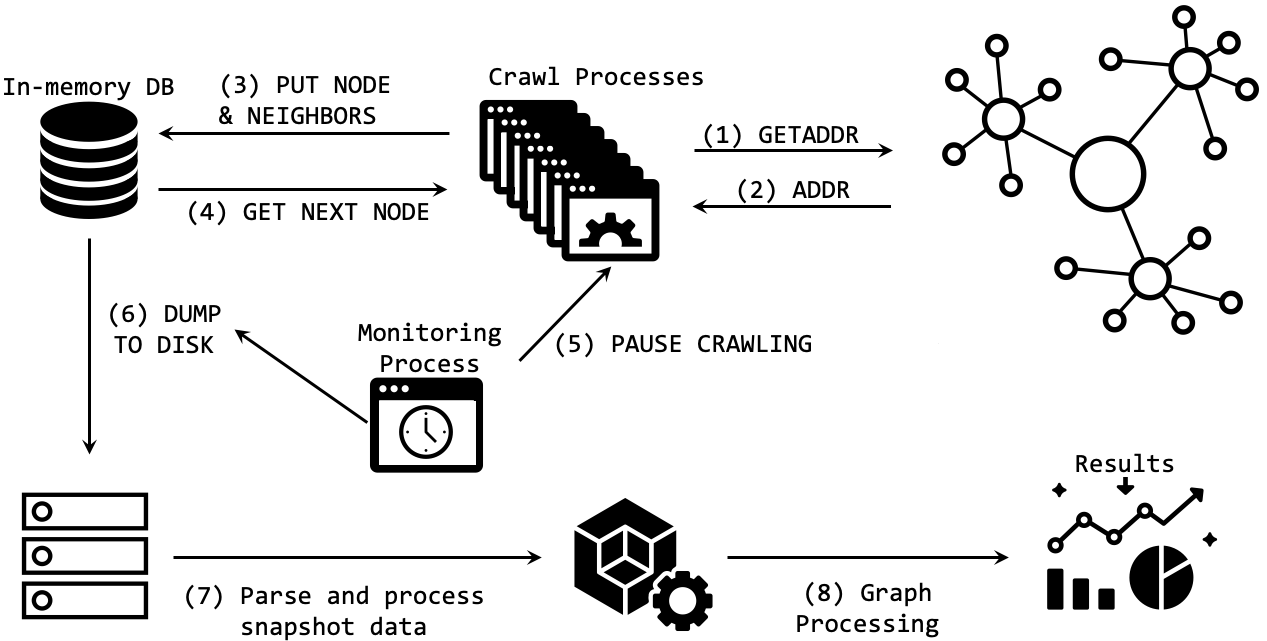}}
  \caption{Data Collection and Processing Pipeline.}
  \label{fig:block-diagram}
\end{figure}

We outline our data collection methodology in Figure~\ref{fig:block-diagram}.
Steps 1-4 correspond to the parallel crawling process.
Steps 5 and 6 represent the synchronized data dump of the collected snapshots from the in-memory database to disk.
In steps 7 and 8 we pre-process and analyze the collected graphs.

\subsection{Crawling Process} \label{sec:methodology-details}
To discover the nodes (peers) of the overlay networks, we modified the crawler maintained by the popular site \emph{Bitnodes.io}~\cite{bitnodes,bitcrawler} to meet our needs. We added features that enable: a) crawling multiple chains using distinct processes; b) storing the mapping of 
each node to its known-peers; c) and synchronizing the processes to dump the collected data for each BC at the same timestamp. 
For Ethereum, which uses a different communication protocol, we built our crawler around the open source Trinity client~\cite{trinity} and disabled all BC-related processing. We aim to scrape the contents of any reachable node’s \texttt{ADDRMAN} for its outgoing connections, and to build a connectivity graph of \textit{any} possible connections that could be realized in the overlay network.

Each BC to be crawled is assigned to a process that creates hundreds of user-level threads. Intermediate data collected during crawling are maintained in an in-memory key-value store, each process having its own  instance. Following the communication protocols of each BC,
each process connects to its assigned network and recursively asks each discovered node for its known peers (\textit{Steps 1-2}). Each new discovered node is stored in a \texttt{pending} set of the in-memory instance (\textit{Step 3}).
The threads constantly poll their \texttt{pending} set for a new node (\textit{Step 4}), initiate a connection and retrieve the list of the node's known peers.

Upon a successful connection to a peer, its entry is moved from \texttt{pending} to the \texttt{tried} set.
On each received reply to a \texttt{getaddr} message, the process makes an entry, mapping the originating node ($N_{or}$) to the peer list it knows of: $N_{or} \rightarrow \{P_0, P_1,...,P_n\} $, where $P_{0-n}$ are the peers included in $N_{or}$'s reply. This entry is stored in the \texttt{edges} set.
When the \texttt{pending} set becomes empty, the crawler moves all entries from \texttt{tried} to the \texttt{pending}, and starts over. The \texttt{edges} set remains intact and is updated in subsequent rounds. 
Replies from nodes that are already mapped in the \texttt{edge} set, are appended to the respective entry.
After a period of approximately two hours, all processes synchronize and dump their \texttt{edge} set to storage (\textit{Steps 5-6}).
Subsequently, after the dump, all sets are emptied and each process restarts and repeats the same procedure.
Note that in our implementation, we do not accept incoming connections and we probe only IPv4 peers.

In Steps 7-8, we construct connectivity graphs using the \texttt{edge} sets collected.
For each entry $N_{or} \rightarrow \{P_0, P_1,\ldots,P_n\} $ in the set,
we create a directional graph with nodes $\{N_{or}, P_0, P_1, \allowbreak \dots, P_n\}$ and add $n$ outgoing edges from $N_{or}$ to nodes $P_0, P_1, \allowbreak \dots, P_n$.
We synthesized and analyzed these graphs using the SNAP~\cite{snap} and NetworkX~\cite{networkx} libraries.

\vspace*{-2mm}
\subsection{Methodology Assessment} \label{sec:methodology-assessment}
We evaluate the efficacy of our method by setting-up an unmodified Bitcoin reference node using the official implementation~\cite{releasenotesbtc}, and probing it using our crawler.
All replies collected by the crawler contain more than 80\% of the client's outgoing peers.
More details can be found in Appendix~\ref{app:methodology-assessment}.


\mysubsubsection{Limitations:}
We readily admit that the advertised peer lists 
may not include all actual connections of a node.  
Furthermore, clients may deviate from the protocol defaults and advertise even less peers. 
Such hidden links cannot be revealed by our method. 
Our resilience assessment relies on the assumption that these links constitute a small minority of all possible links.


\mysubsubsection{Note:}
\label{sec:bitcoin-topology-hiding}
The latest version of Bitcoin Core includes changes that affect the proposed approach.
This version employs cached responses to \texttt{getadrr} requests; for a period of 24 hours, all \texttt{getadrr} requests from any peer are served by a fixed \texttt{addr} reply containing up to 1000 peers.
This further impedes the collection of contents of \texttt{ADDRMAN}~\cite{pr18991}. 
If adoption of the latest client version prevails, we could expect a reduction in both size and density of Bitcoin's synthesized graphs, narrowing the view extracted by our proposed methods. 
Nevertheless, a motivated adversary can still infer most critical nodes based on other metrics (see Sec.~\ref{ssec:results-uptime-robust}).

\subsection{Datasets \& Experiments} \label{sec:expsetup}
Using the aforementioned methodology, we crawled the selected BC networks from the datacenter of a European University.
The monitoring server has an 8-core/3.2GHz CPU, 64GB RAM, and 2.1TB of HDD storage. The crawling operations were done for a period of about one month (26/06-22/07/2020). At the end of the crawling period, we had collected 335 network snapshots for each BC network, or 2345 graphs in total. 
The collected dataset and the analysis scripts are anonymously available for review at~\cite{review-dataset-code-link}. 
Our ethical considerations are outlined in Appendix~\ref{app:ethics}.

We denote as $C$ the set of the 7 BC networks crawled. 
At the end of every two-hour period, we have seven different \texttt{edge} sets, one per BC $c \in C$. All such edge sets are annotated with the timestamp $t$ of their crawl. Each set of edges corresponds to a graph, denoted as $S_{c}^{t}$, representing a snapshot of BC network $c$, at timestamp $t$.

\section{Network Structure}
\label{sec:results}
\label{sec:results-net}

This section is driven by Research Question 1. In particular, we are interested in answering the following questions: a) What are the structural properties and network characteristics of BC overlay networks? b) Are they all structured in a similar manner? c) Do they share common properties? d) Do they have properties that relate to other well-known networks, 
or do they resemble random networks?

\subsection{Basic network properties}
\begin{table}[t]
  \centering
  \caption{Basic network graph metrics per BC network (average values across all collected snapshots.)}
  \resizebox{1.\columnwidth}{!}{%
    \begin{tabular}{|p{0.35\linewidth}|r|r|r|r|r|r|r|}
  \hline
    	Network:&
      	Bitcoin&
		Bitcoin Cash&
		Dash&
		Dogecoin &
      	Ethereum &
		Litecoin &
		Zcash
      \\
  \hline
    Nodes &
		50k &
		23k &
		8.5k &
		1.2k &
		12k &
		8.2k &
		1.5k 
      \\
  \hline
    Edges &
      4794k  &
      169k  &
      7312k  &
      116k  &
      59k  &
      741k  &
      106k 
      \\
  \hline
    Connected Component &
      0.99 &
      0.99 &
      1 &
      1 &
      0.99 &
      1 &
      1
      \\
  \hline
  Strognly Connected Component &
      0.11 &
      0.04 &
      0.82 &
      0.27 &
      0.04 &
      0.16 &
      0.16
      \\
      \hline
    Diameter &
      4 &
      4 &
      4 &
      3 &
      6 &
      4 &
      4
      \\
      \hline
    Density &
      0.002 &
      3E-04 &
      0.1035 &
      0.0805 &
      0.0006 &
      0.0112 &
      0.0617
      \\
      \hline
    Avg. Degree &
      92.8 &
      7.12 &
      802.12 &
      74.97 &
      4.3 &
      104.04 &
      56.96
      \\
      \hline
    Assortativity &
      -0.2 &
      -0.64 &
      -0.06 &
      -0.13 &
      -0.02 &
      -0.01 &
      -0.22
      \\
      \hline
    Reciprocity &
      -0.06 &
      0.05 &
      0.16 &
      0.34 &
      0 &
      0.09 &
      0.47
      \\
      \hline
    Global Clustering Coefficient &
      0.049 &
      0.011 &
      0.166 &
      0.28685 &
      0.0022 &
      0.0735 &
      0.3094
      \\
      \hline
    Avg. Shortest Path &
      2.55 &
      2.82 &
      1.93 &
      1.77 &
      3.78 &
      1.96 &
      1.72
      \\
  \hline
    \end{tabular}%
  }
  \label{tab:all-metrics}%
  \vspace{-5mm}
\end{table}%

The basic properties of the derived graphs are summarized in Table~\ref{tab:all-metrics}. The metrics were computed individually on each graph $S_{c}^t$ and were then averaged.
All networks appear to be well--connected given the size of their largest connected component and low diameters. 
Moreover, we observe that Dash is markedly the most dense network and is almost fully connected. It has a strongly connected component (SCC), \ie a subgraph in which every node is reachable from every other node. The SCC comprises 82\% of the total network nodes. Large BC networks have a smaller SCC compared to the smaller ones. 
Values extracted from our datasets match reported values in related measurement works~\cite{Kim18,btcmap,mapz}. 
Indicatively, on each day, our monitoring node was able to discover $120081$ nodes in Bitcoin, $19543$ in Ethereum, and $4132$ in Zcash (reporting median values).
On average, the monitoring node performed more than $1.3M$ requests per day, covering all BC networks.

\subsection{Degree Distributions}
\begin{figure*}
\centering
    \subfloat[\centering Bitcoin]{{\includegraphics[width=0.3\linewidth]{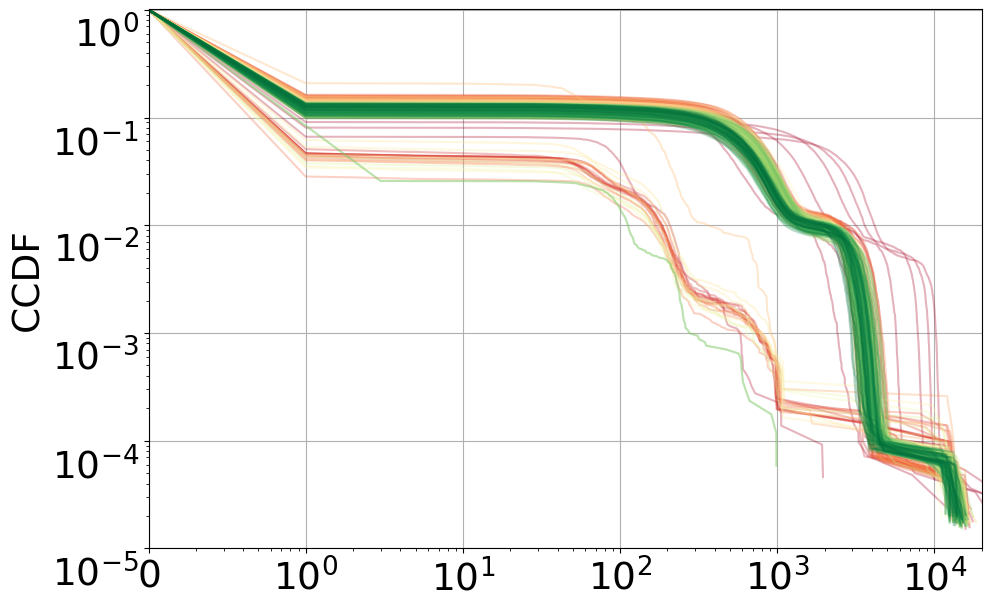}}}%
    \subfloat[\centering Bitcoin Cash]{{\includegraphics[width=0.3\linewidth]{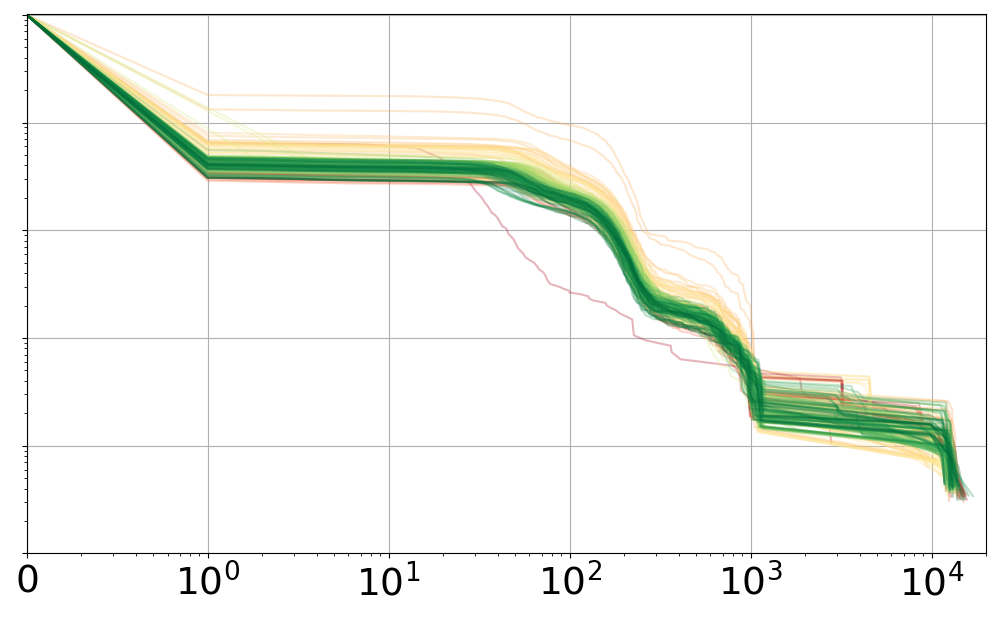}}}%
    \subfloat[\centering Dash]{{\includegraphics[width=0.3\linewidth]{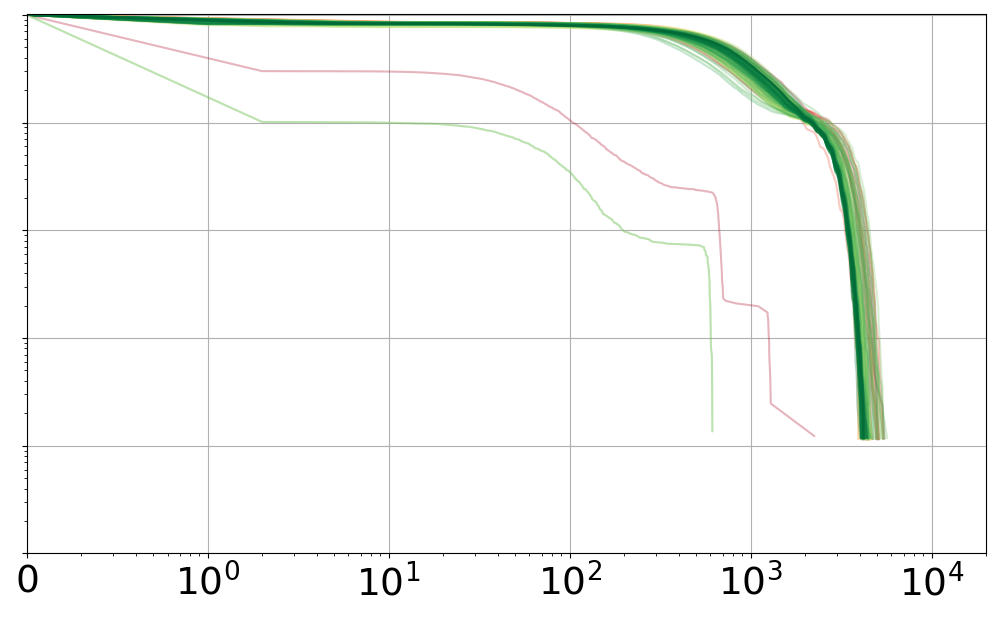}}}%
    \subfloat[\centering Colorbar]{{\includegraphics[width=0.105\linewidth]{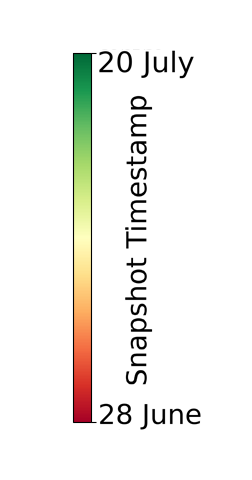}}}%
    \:
    \subfloat[\centering Dogecoin]{{\includegraphics[width=0.25\linewidth]{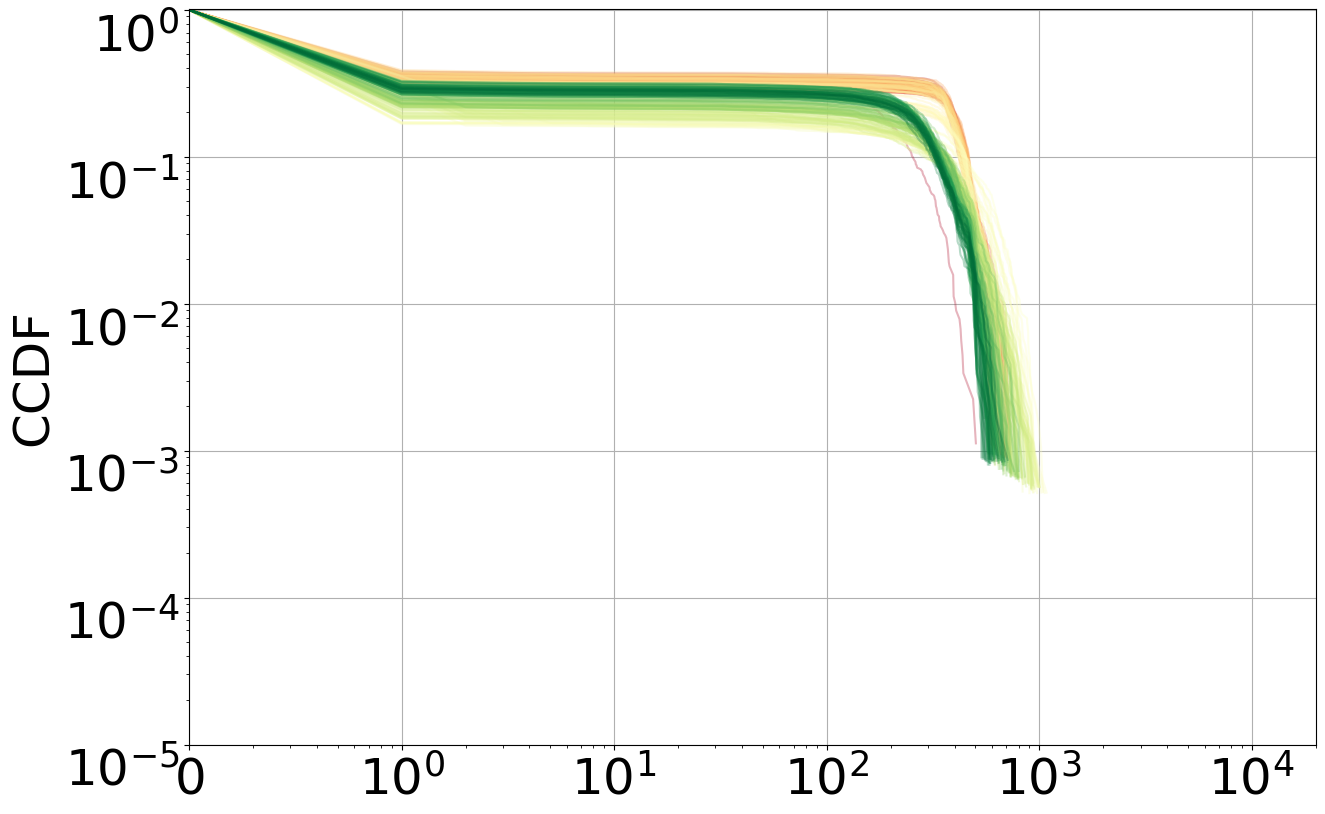}}}%
    \subfloat[\centering Ethereum]{{\includegraphics[width=0.25\linewidth]{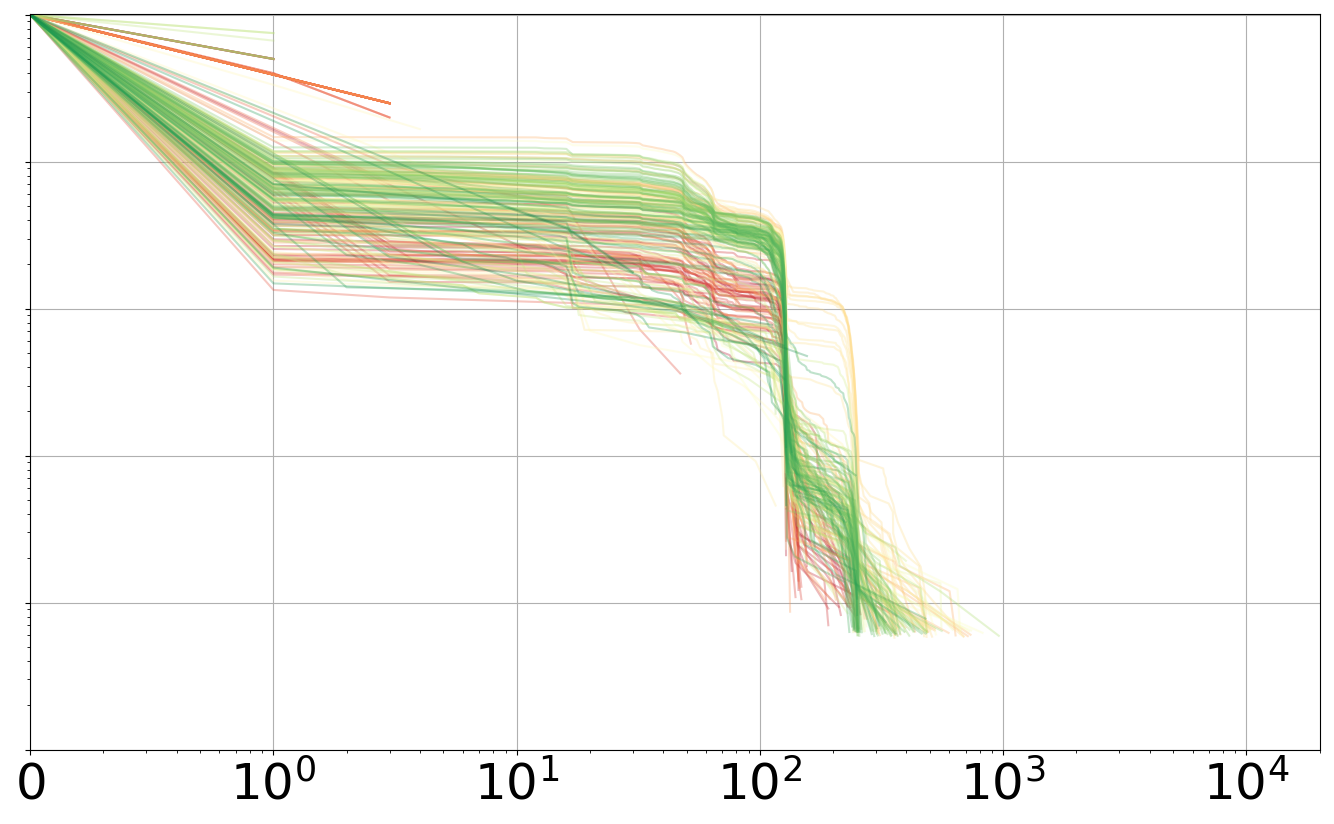}}}%
    \subfloat[\centering Litecoin]{{\includegraphics[width=0.25\linewidth]{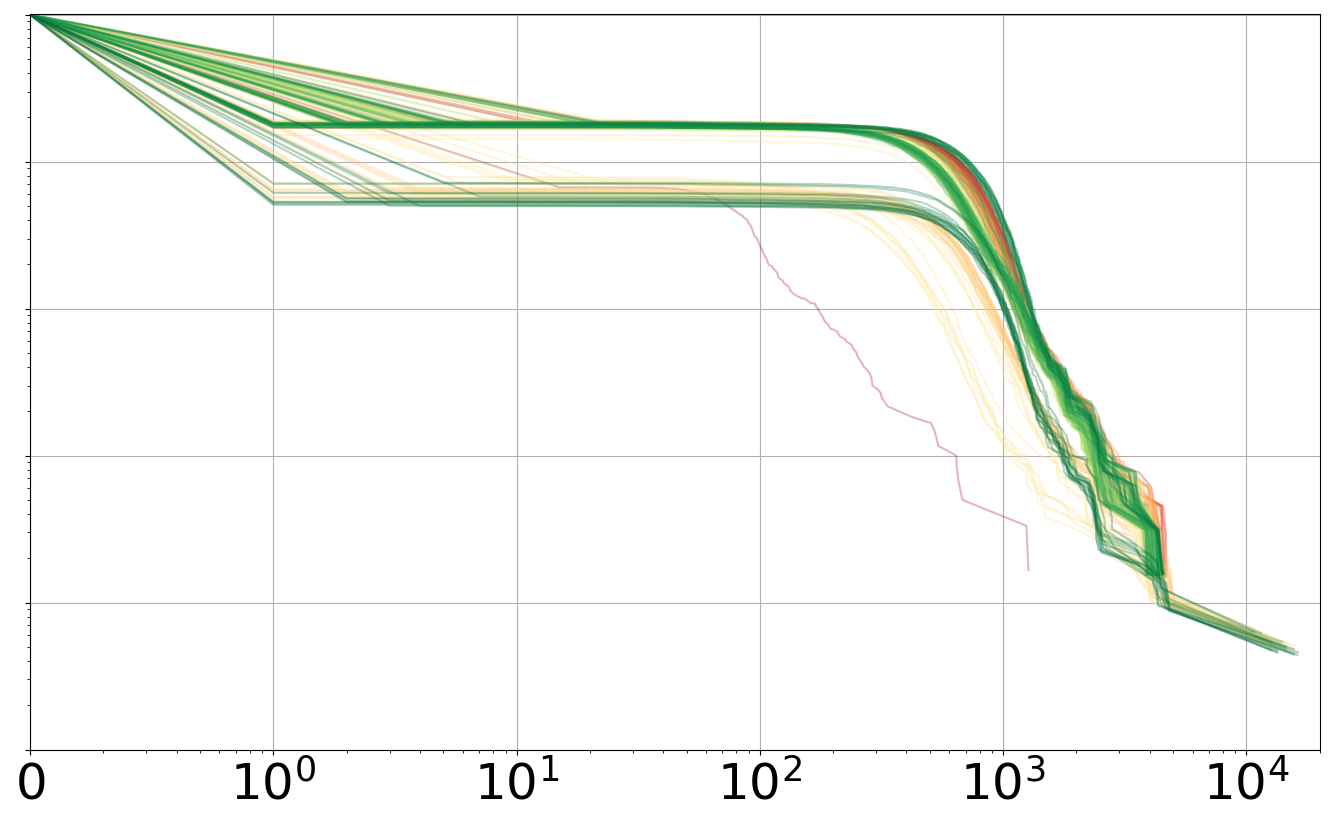}}}%
    \subfloat[\centering ZCash]{{\includegraphics[width=0.25\linewidth]{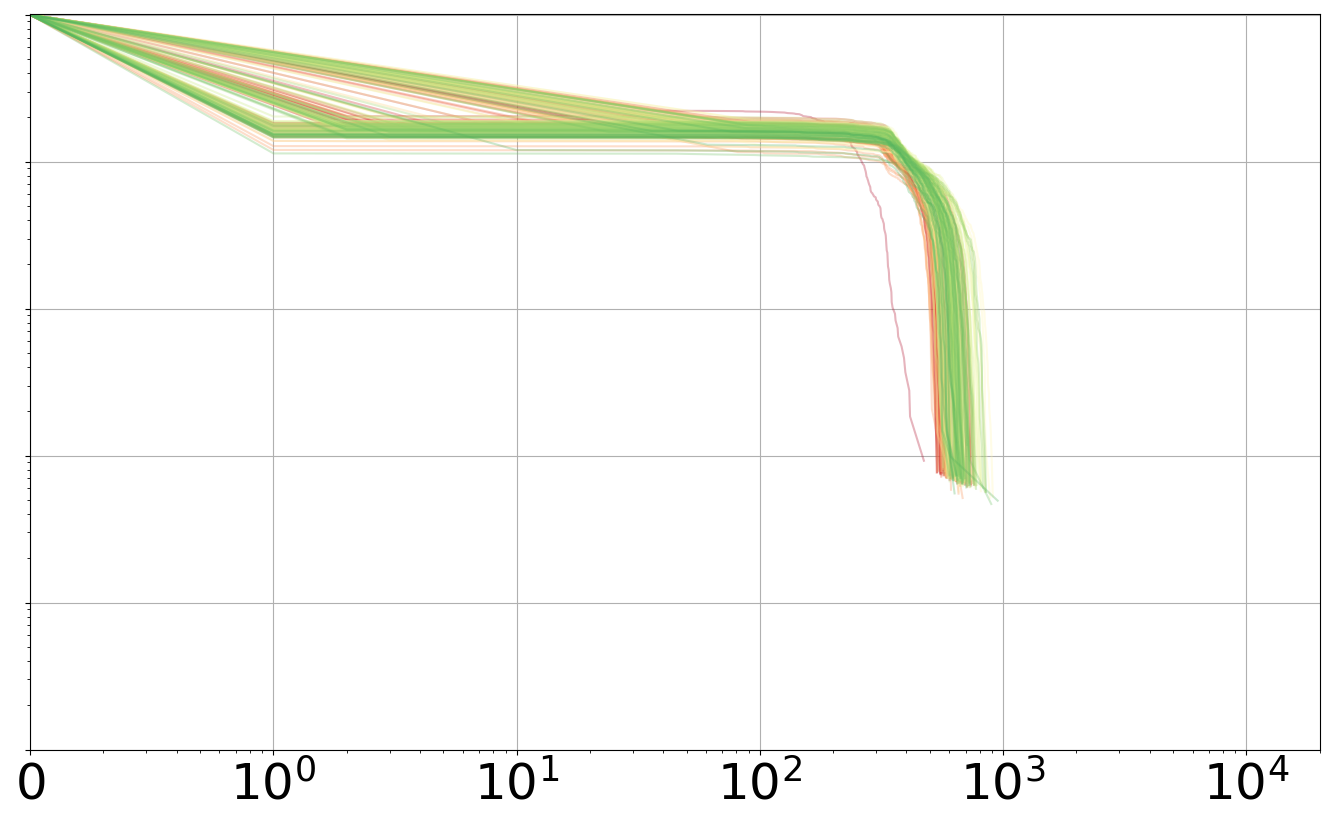}}}%
    \caption{Out-degree CCDF of BC networks studied.
    Snapshots are colored according to the colorbar. 
    }%
\label{fig:deg-dist}%
\end{figure*}
The degree (number of links with other nodes) distribution affects many network phenomena, like network robustness and efficiency in information dissemination~\cite{barabasi2016network}.
In addition, random networks have binomial degree distributions, while in real systems we usually encounter highly connected nodes that the random network model can not account for.

In Figure~\ref{fig:deg-dist}, we plot the CCDF of the out-degree of all collected snapshots for all BC networks studied.
\hyphenation{Eth-ereum}
We color the snapshots according to their timestamp. Our first observation is that networks such as Bitcoin and Ethereum manifest considerable variability in degree distribution between snapshots. On the contrary, degree distributions in Dash and Dogecoin have less variability (seen by the distance between snapshots). Another interesting observation is that in most BC networks, we have a high fraction of unreachable nodes, either because they are offline or behind NATs. This is indicated by the high fraction of nodes with zero out-degree.
This observation confirms the findings of Wang and Pustogarov~\cite{WangPustogarov17} who studied the prevalence and deanonymization of unreachable peers. 
Our results also suggest that these BC network have heavy-tailed degree distributions.
We further discuss their best distribution fit and their scale--free property in a following paragraph.
Finally, we observe significant deviations from the BC protocols. 
In Bitcoin for instance, one would expect that reachable nodes would have at least 1K out-degree, since Bitcoin clients with the default parameters are set to respond with 1K known peers.
Conversely, we observe a number of nodes with out-degree less than 100.

\subsection{Scale-free property}\label{sub:scalefree}
One network property tightly related with the degree distribution of a network is the scale-free property.
A scale-free network is defined as a network whose degree distribution follows a power-law, \ie having a probability distribution $p(k) \propto k^{-\alpha}$. Exponent $\alpha$ is known as the scaling parameter, and typically lies in the range $2<\alpha<3$.
The scale-free property strongly correlates with the network's robustness to random failures and has received tremendous attention in the scientific literature (\eg~see~\cite{barabasi2016network}).
Many real-world networks have been reported to be scale-free, although their prevalence is questioned~\cite{clauset}.

To test how well the degree distribution of each BC network snapshot can be modeled by a \emph{power-law} ($PL$), \emph{log-normal} ($LN$), \emph{power-law with exponential cutoff} ($PLEC$) or \emph{stretched exponential} ($SE$), we calculate the best fit using the \emph{powerlaw} package available by Alstott \etal~\cite{Alstott_2014}.
\begin{table}[h]
  \caption{Degree distributions of BC networks best-fit for different types of exponential distributions.\\
$PL$: power-law;
$LN$: log-normal;
$PLEC$: power-law with exponential cutoff;
$SE$: stretched exponential.}
\label{tab:powerlaw}
    \begin{tabular}{|l|r|r|l|r|}
    \hline
    \textbf{Disrtibution} &
      \multicolumn{1}{l|}{\textbf{LN}} &
      \multicolumn{1}{l|}{\textbf{PL}} &
      \textbf{PLEC} &
      \multicolumn{1}{l|}{\textbf{SE}}
      \\
    \hline
    Bitcoin &
      6.29\% &
      0.60\% &
      \multicolumn{1}{r|}{93.11\%} &
      \multicolumn{1}{l|}{-}
      \\
    \hline
    \multicolumn{1}{|l|}{Bitcoin Cash} &
      76.90\% &
      16.20\% &
      \multicolumn{1}{r|}{6.90\%} &
      \multicolumn{1}{l|}{-}
      \\
    \hline
    Dash &
      \multicolumn{1}{l|}{-} &
      1.80\% &
      \multicolumn{1}{r|}{57.20\%} &
      41\%
      \\
    \hline
    Dogecoin &
      49.40\% &
      4.80\% &
      - &
      45.80\%
      \\
    \hline
    Ethereum &
      21.90\% &
      24.60\% &
      \multicolumn{1}{r|}{18.30\%} &
      35.30\%
      \\
    \hline
    Litecoin &
      40.10\% &
      12.60\% &
      \multicolumn{1}{r|}{46.40\%} &
      0.90\%
      \\
    \hline
    Zcash &
      0.60\% &
      18.90\% &
      - &
      80.50\%
      \\
    \hline
    \end{tabular}%
\end{table}%
In Table~\ref{tab:powerlaw}, we report the number of times each type of distribution was the best fit, for all snapshots of the same BC network. 
The calculated results indicate the dynamic nature of BC networks.
Such BC networks that change over time may fit different distributions depending on the snapshot collected, something that is also visible in Figure~\ref{fig:deg-dist}. The results suggest that BC networks are not structured in the same way.
Nevertheless, their degree distributions, in general, belong to the exponential family.
According to sources~\cite{txprobe,Decker2013, pingoletSOK}, Bitcoin’s network formation process is intended to induce a random graph.
Surprisingly, we find that BC networks are different than random networks, confirming past studies~\cite{coinscope, txprobe}.
\subsection{Degree Assortativity}
In general, a network displays degree correlations if the number of links between the high and low-degree nodes is systematically different from what is expected by chance.
In some types of networks, high-degree nodes (or hubs) tend to link to other such hubs, while in other types, hubs tend to link to low-degree nodes, \ie~what is known as a hub-and-spoke pattern.
Assortativity, or assortative mixing is a preference for nodes in a network to attach to others that are similar in some property; usually a node's degree. 
Correlations between nodes of similar degree are often found in the mixing patterns of many observable networks.  For instance, social networks tend to be assortative, while technological and biological networks typically show disassortative mixing, as high-degree nodes tend to attach to low-degree nodes.

We compute the assortativity coefficient for each snapshot of BC network and report the average value over all snapshots in Table~\ref{tab:all-metrics}. We find that Dash, Ethereum, and Litecoin have neutral assortativities. Conversely, Bitcoin Cash, Zcash, and Bitcoin are more dissasortative.
Negative assortativity reveals a hub-and-spoke network structure, and hints to the existence of central, \ie~important network peers.

\vspace*{-3mm}
\subsection{Out/In Degree Ratio}
\begin{figure*}[h]
\centering
    \subfloat[\centering Bitcoin]{{\includegraphics[width=0.33\linewidth]
    {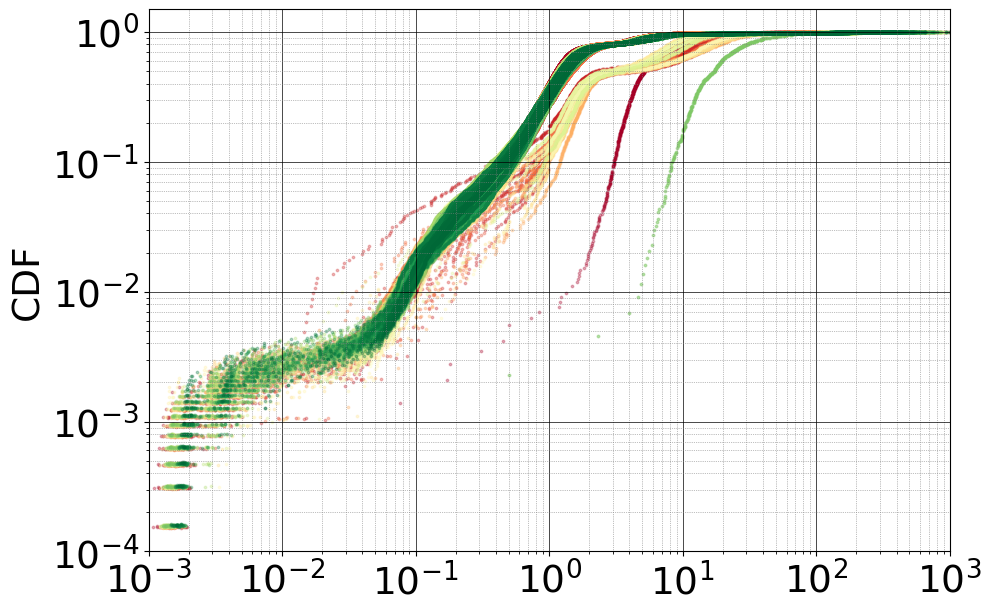}}}%
    \subfloat[\centering Bitcoin Cash]{{\includegraphics[width=0.33\linewidth]
    {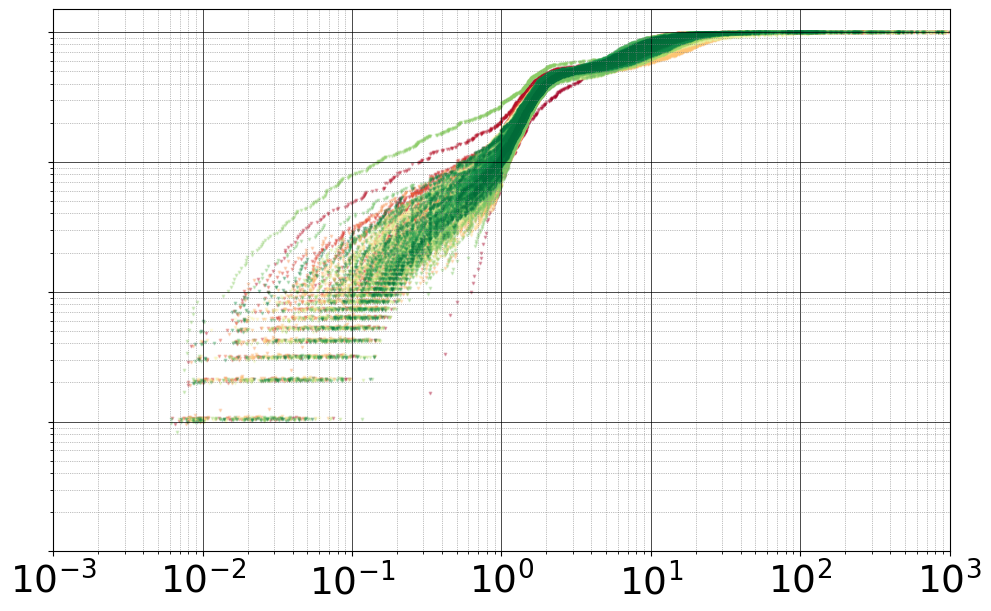}}}%
    \subfloat[\centering Dash]{{\includegraphics[width=0.33\linewidth]
    {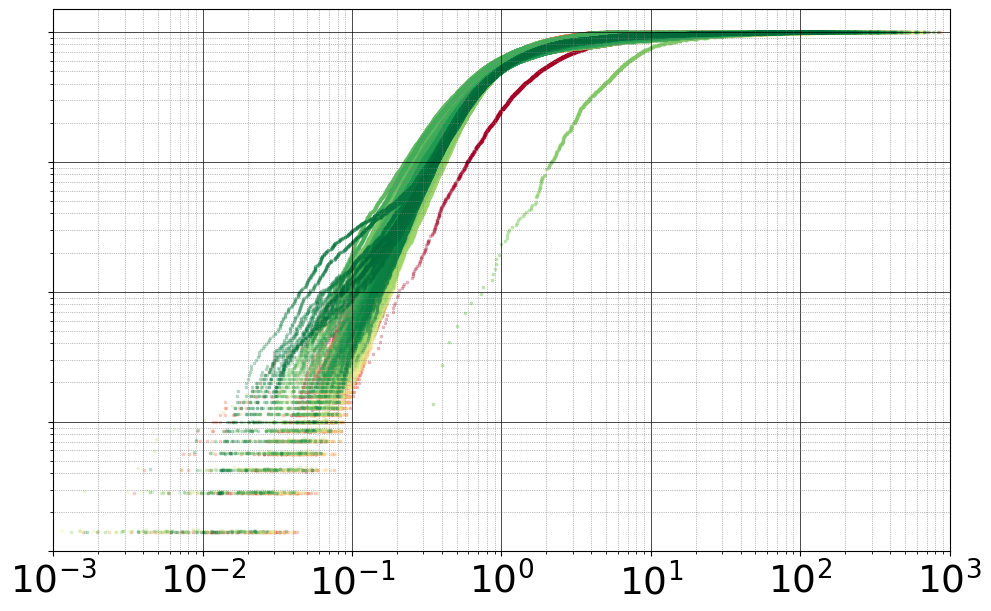}}}%
    \:
    \subfloat[\centering Dogecoin]{{\includegraphics[width=0.25\linewidth]
    {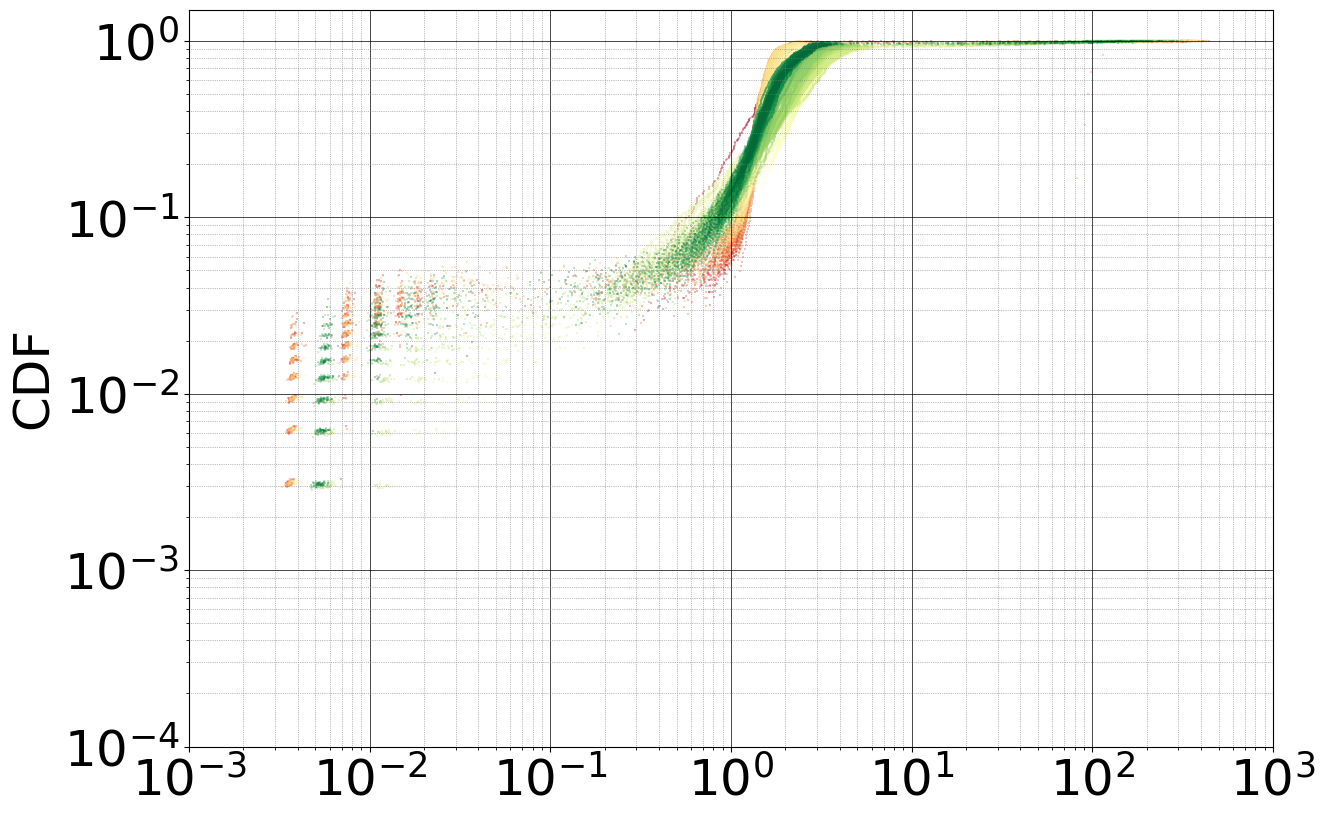}}}%
    \subfloat[\centering Ethereum]{{\includegraphics[width=0.25\linewidth]
    {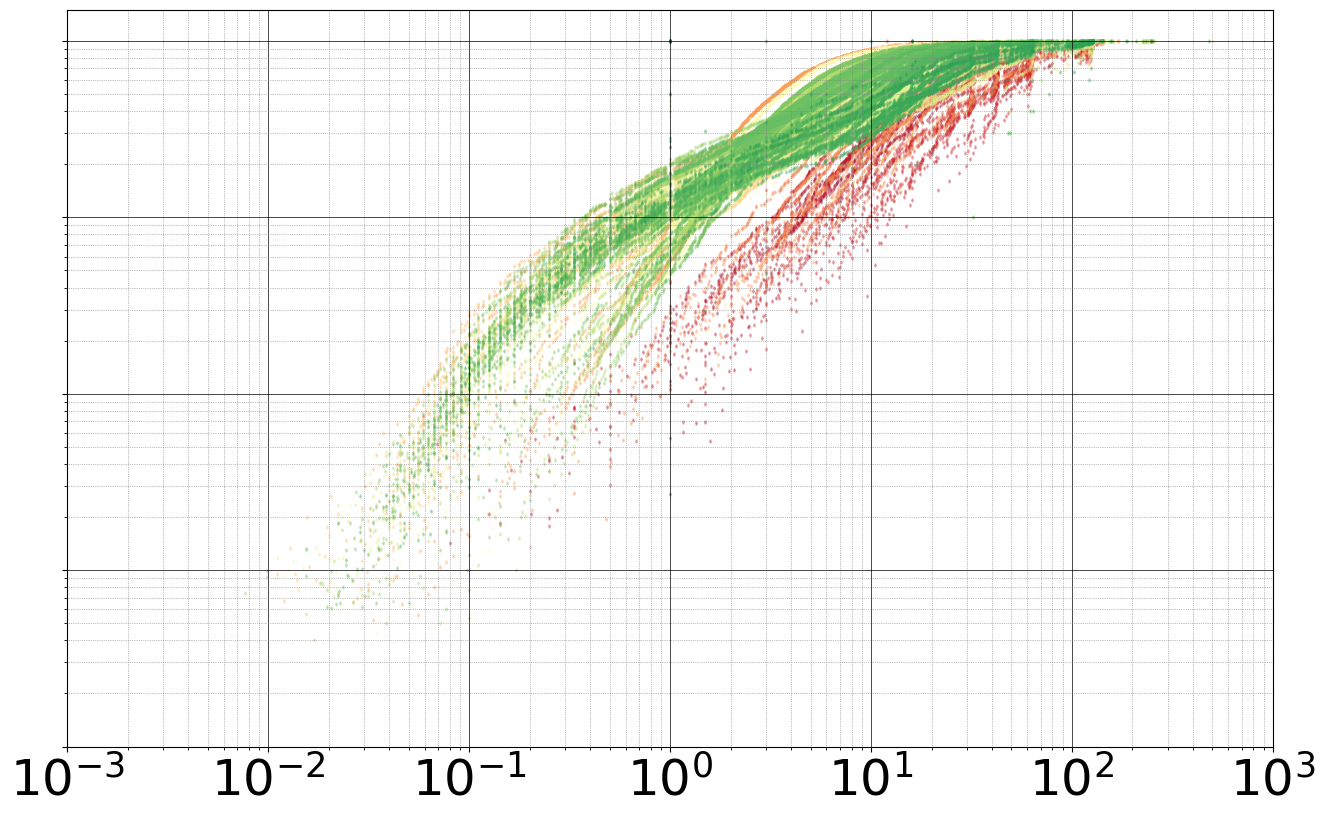}}}%
    \subfloat[\centering Litecoin]{{\includegraphics[width=0.25\linewidth]
    {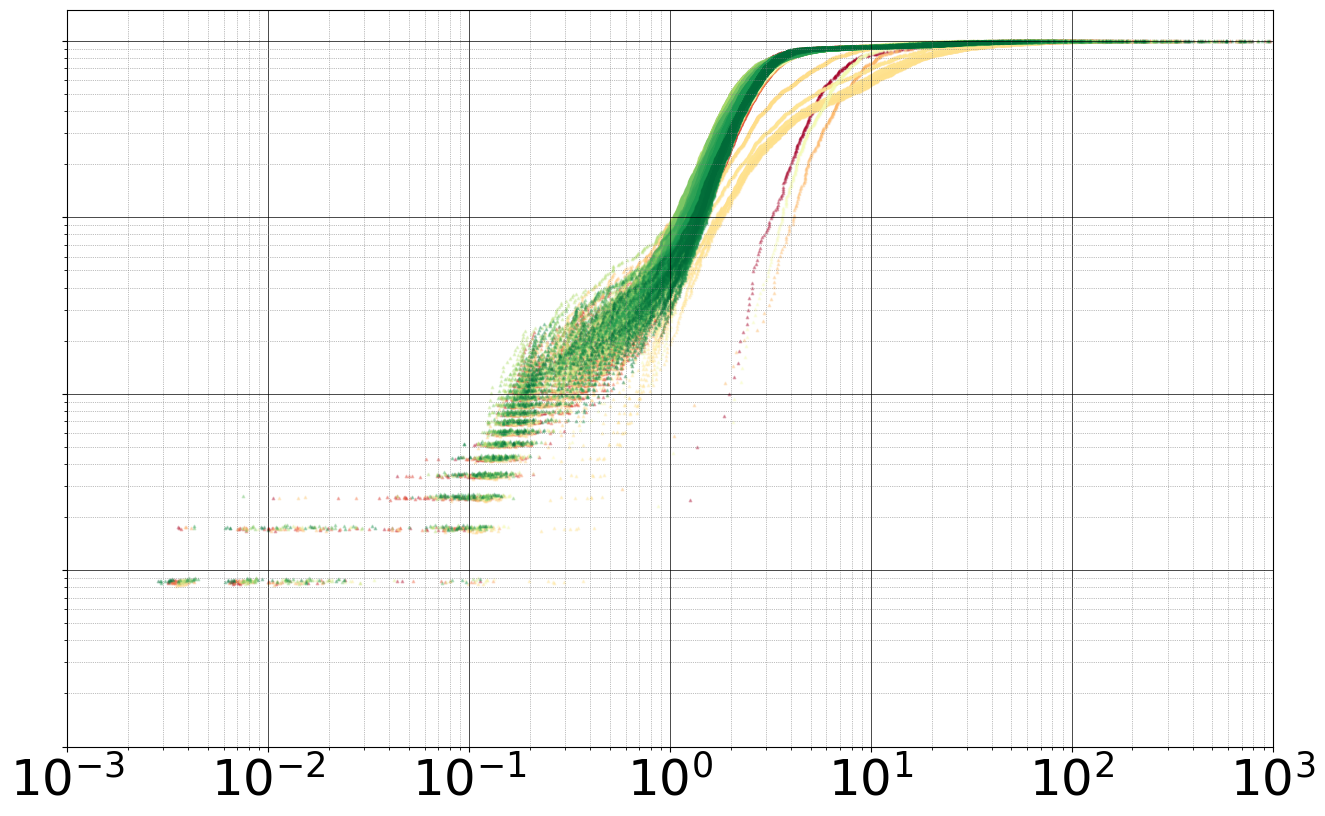}}}%
    \subfloat[\centering Zcash]{{\includegraphics[width=0.25\linewidth]
    {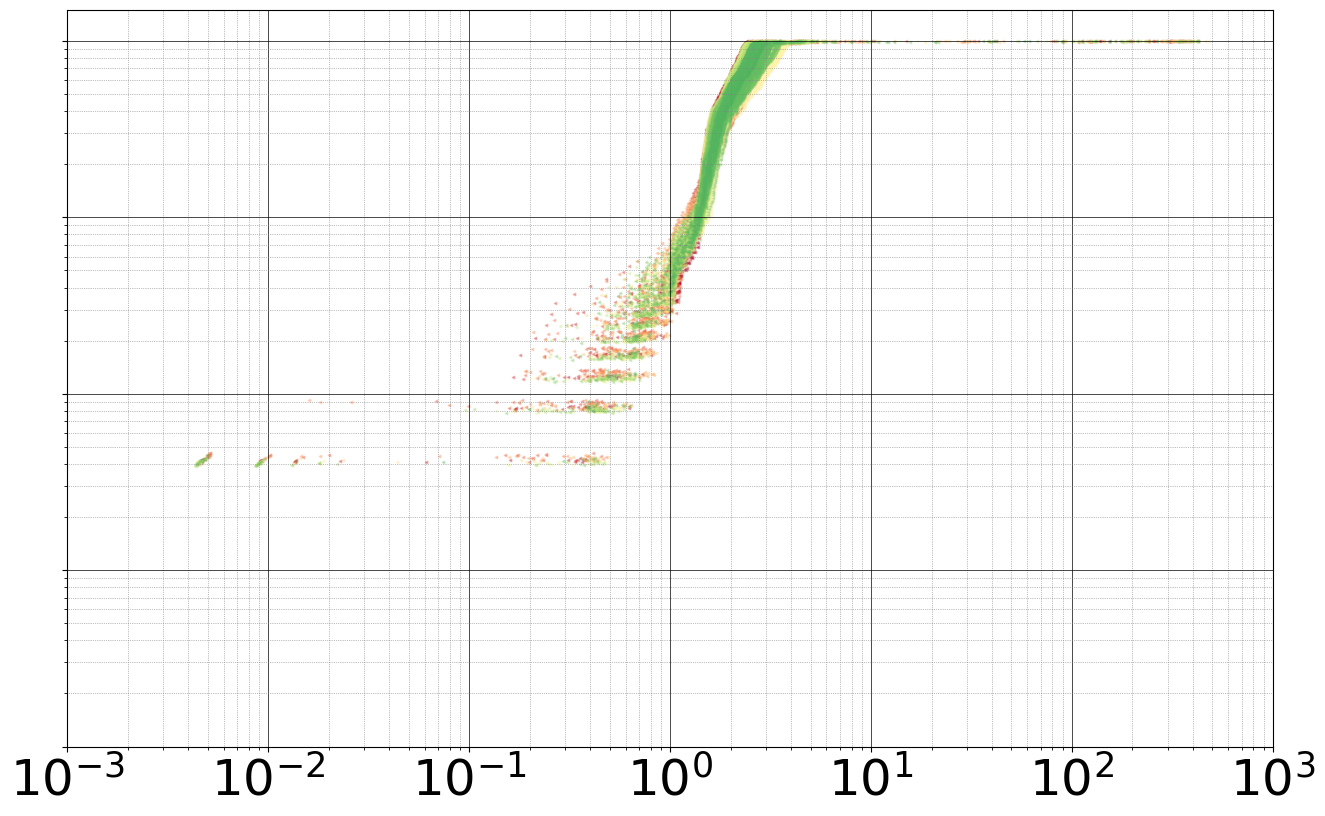}}}%
    \caption{CDF of $Out/In$ ratio per BC network studied.
    Snapshots are colored according to Figure~\ref{fig:deg-dist}.
    }%
\label{fig:deg-corr-dist}%
\end{figure*}
Link analysis using the in-degree and out-degree distributions
has proven very powerful in identifying authoritative and central nodes in the Web and social networks ~\cite{webstruct}. We computed the out-degree over in-degree ratio of individual nodes for all snapshots and use it to compare the structure of BC networks to the Web and social networks.
In the Web, most nodes have considerably higher out-degrees than in-degrees ($\frac{Out}{In} > 1$), 
while a small fraction of nodes have significantly higher in-degrees than out-degrees ($\frac{Out}{In}
< 1$). Social networks have substantial correlation between in-degree and out-degree and most nodes 
have an in-degree within 20\% of their out-degree~\cite{socialnetsanalysis}.
In Figure~\ref{fig:deg-corr-dist}, we show the CDFs of the outdegree-to-indegree 
ratio for nodes in all networks. Dogecoin marginally resembles a social network with 45\% of its
nodes having a good correlation, within 20\%. This can be explained by the high number of symmetric 
links, since Dogecoin has high reciprocity (see Table~\ref{tab:all-metrics}).
In Bitcoin and Bitcoin Cash, most nodes have considerably higher out-degree than in-degree, while a small fraction of nodes have significantly higher in-degree than out-degree, a characteristic similar to the Web.
Their resemblance to the Web can be further strengthened by their dissassortative nature (see Table~\ref{tab:all-metrics}).
In Dash, we cannot observe any correlations, with half the nodes 
having higher out-degree and the other half having higher in-degree.
This indicates that half of peers in Dash are frequently online in contrast with the other half that participate periodically.
Our longitudinal analysis in Sec.~\ref{sec:results-temp}, confirms this indication.
In Litecoin and Zcash, we observe a concentration of nodes having  higher out-degrees, and compared with their assortativity, this indicates that nodes tend to connect to lower degree nodes. 
In Ethereum, due to its distinguished discovery mechanism, the vast majority of nodes have a very high out- over in-degree ratio.

\subsection{Reciprocity}
The reciprocity property is a measure of likelihood of vertices in a directed graph to be mutually linked. It has been shown to be critical in modeling and classification of directed networks~\cite{Garlaschelli_2004}. Table~\ref{tab:all-metrics} lists the average reciprocity across all snapshots for each BC network. Zcash, Dogecoin, and Dash have significantly higher reciprocity values. We can attribute this finding to a similar explanation on size and clustering.
Closer examination also reveals that these networks have a considerable number of nodes that are frequently online: such nodes are more likely to connect to nodes also frequently connected, driving reciprocity higher.

\vspace*{-4mm}
\subsection{Clustering Coefficient ($CC$)}
The local clustering coefficient $CC_i$ measures the density of links in node $i$’s immediate neighborhood:
$CC_i = 0 $ means that there are no links between $i$’s neighbors, while
$CC_i = 1 $ implies that each of $i$’s neighbors link to each other as well.
In a random network, the local $CC$ is independent of the node’s degree, and average $CC$, \ie~$\overline{CC}$, depends on the system's size with respect to nodes, $N$. In contrast, measurements indicate that for real networks, \eg~the Internet, the Web, Science collaboration networks, 
the $CC$ decreases with the node's degree and is largely independent of the system size~\cite{barabasi2016network}.
The local $CC$ in a random network ($CC_{rand}$) is calculated as the average degree $\overline{k}$ over $N$, \ie~$CC_{rand} = \frac{\overline{k}}{N}$.
The average degree of a network is equal to $\frac{2L}{N}$, where $L$ is the number of links.
The $\overline{CC}$ of a real network is expected to be much higher than that of a random graph.
For the clustering coefficient we consider undirected versions of the graphs.
\begin{figure}[h]
    \centering
    \subfloat[\centering All Networks]{
    {\includegraphics[width=0.5\linewidth]{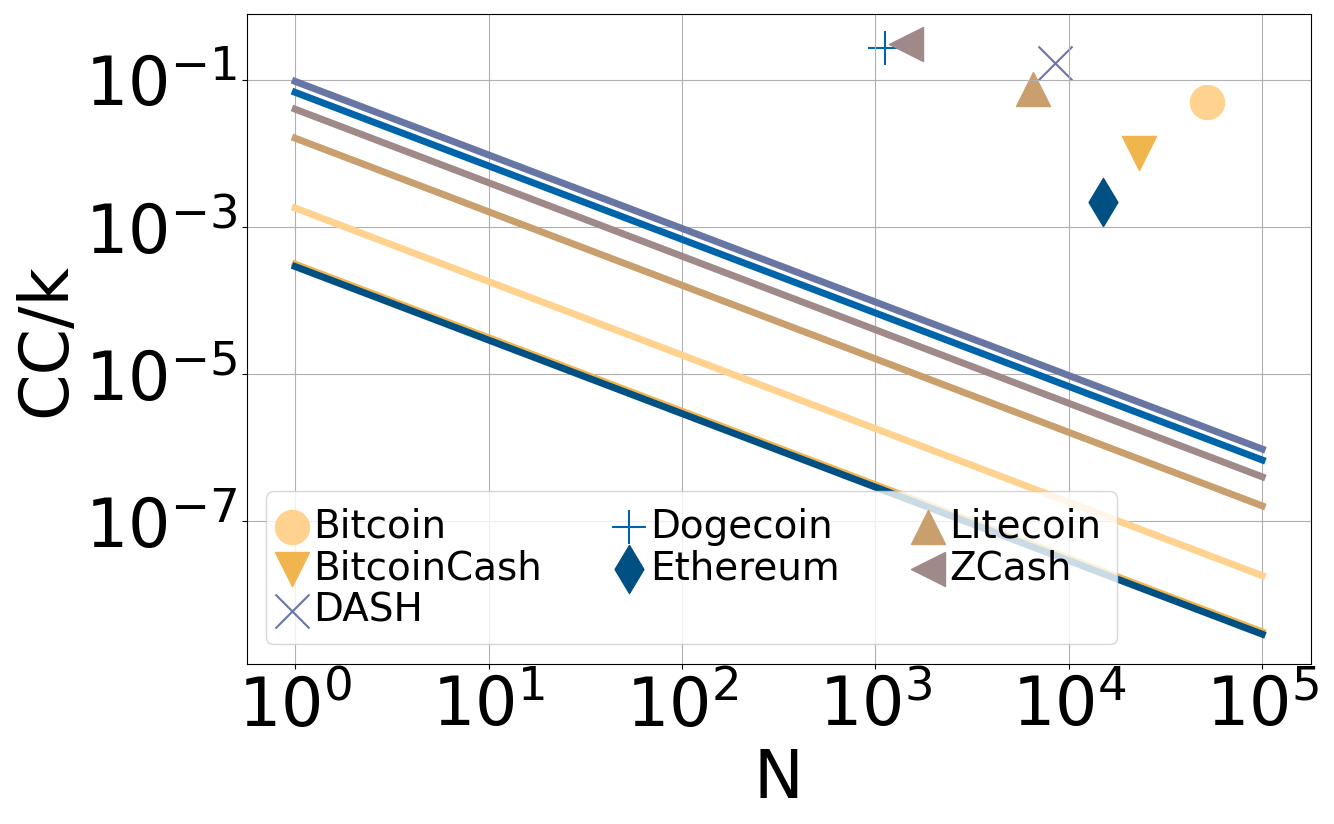}}
    }%
    \subfloat[\centering Bitcoin and Dash]{
    {\includegraphics[width=0.5\linewidth]{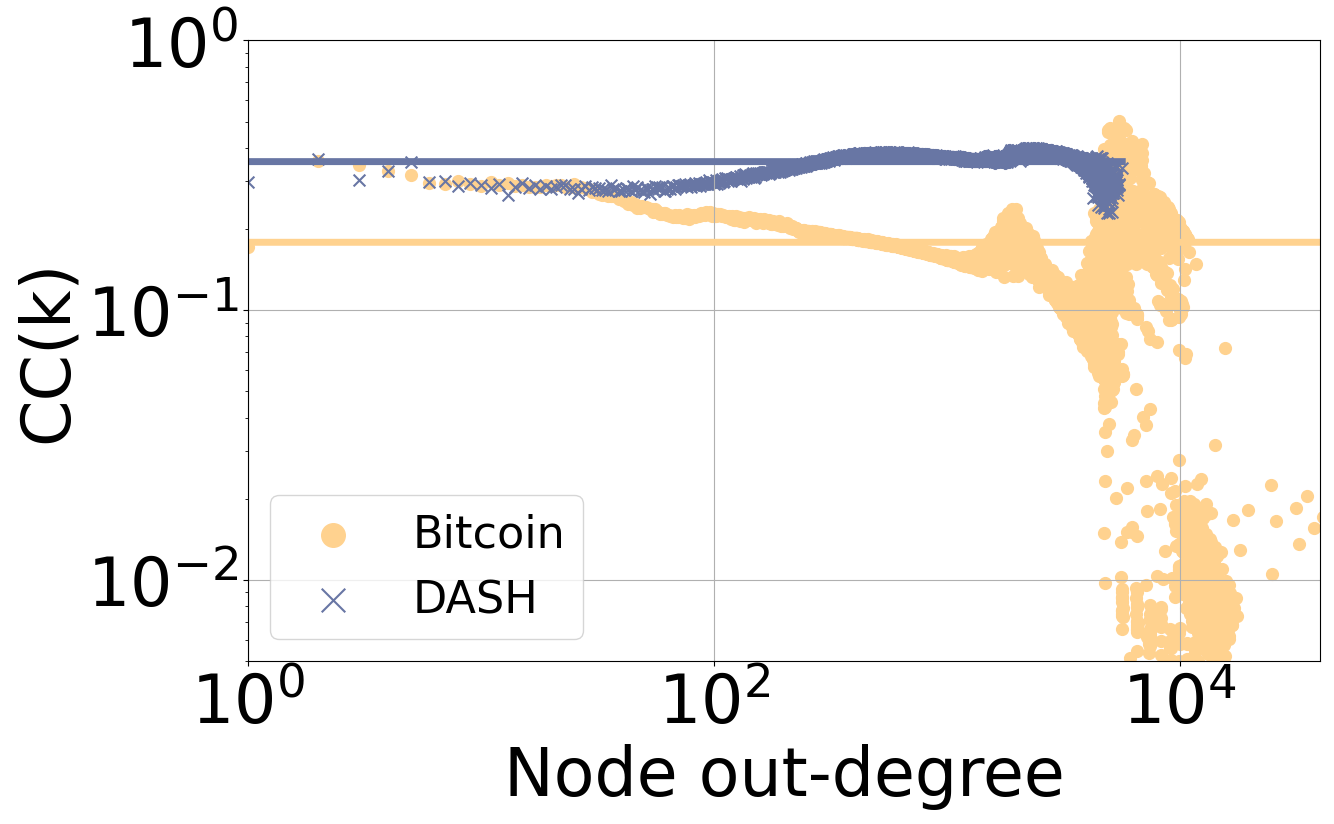}}
    }%
    \caption{Analysis of Clustering Coefficient (CC) results:
    (a) $\frac{\overline{CC}}{\overline{k}}$ vs. BC network size; Size and CC averaged across snapshots $S_{c}^{t} \forall t \in T$.
    Markers correspond to the networks of Table~\ref{tab:all-metrics}. 
    Lines correspond to the prediction for random networks, $CC = \frac{\overline{k}}{N}$, with constant $\overline{k}$ and varying size $N$. 
    Similar to other known networks, the average $CC$ appears to be independent of the network size $N$.
    (b) The dependence of the local $CC$ on the node’s degree for each network.
    Y-axis represents $CC(k)$, measured by averaging the local $CC$ of all nodes with the same degree $k$ (showing results of aggregating all snapshots of a given BC network).
    Horizontal lines correspond to the global $CC$ of the network. 
    Other BC networks are omitted for clarity.}
    \label{fig:cc-all}
    \vspace*{-3mm}
\end{figure}
In Figure~\ref{fig:cc-all}(a), we compare the $\overline{CC}$ of the BC networks with the expected $CC$ for random networks 
of similar size. As in other real networks, we observe higher $CC$ than expected for a random network, further indicating 
that BC networks deviate significantly from random networks. In Figure~\ref{fig:cc-all}(b), we plot the dependence of
the $CC$ on the node's degree for two of the BC networks under study, where we make some remarkable observations. 
Although the empirical rule from~\cite{barabasi2016network} states that higher degree nodes have lower $CC$, in Bitcoin we observe a significant fraction of high degree nodes with high $CC$. The same finding was observed in Ethereum and Zcash 
networks as well.
Another deviation from the same empirical rule is observed in Dash, where all nodes appear to have an almost constant 
$CC$, independent of node degree. We attribute this behavior to its temporal characteristics, previously discussed in results related to Fig.~\ref{fig:deg-dist}. These are further studied in Sec.~\ref{sec:results-temp}. The observed $CC$ distributions indicate that BC networks are, in general, governed by rules rarely encountered in other known networked systems.

\subsection{Small-world Property}
We did not find evidence that the networks under study satisfy the small-world property.
Although we observe low average distances in all BC networks, they do not have high enough clustering coefficients to be considered as small-world. More details are discussed in Appendix~\ref{app:small-world}.

\section{Overlaps between BC Networks} \label{sec:results-overlaps}
In this section, we address the second research question \textit{RQ2}. 
We define as \textit{overlapping nodes} those nodes that participate in more than one network at the same timestamp.
Nodes are identified by their IP address. 
The intuition of our analysis is as follows.
In each snapshot, we compare the set of \textit{overlapping} nodes with all 
the other nodes, in order to draw insights on \textit{overlapping} nodes' properties.
%
Before describing the details of our study, we outline our mathematical notation to help explain our analysis.
As mentioned earlier, the set of BCs studied is $C$.
A snapshot of BC network $c$, at timestamp $t$ is denoted as $S_{c}^{t}$.
We define the set $\textbf{S}$ as our collected data-set, that consists of all snapshots $S_{c}^t$.
We denote as $S_{C}^t$ the subset of $\textbf{S}$ that contains all networks at time $t$.
Subsequently, for each snapshot $S_{c}^{t} \in S_{C}^t$ we define two groups, 
$G_{c}^{t}$ and $G_{c}^{'t}$, such that 
$ G_{c}^{t} = S_{c}^{t} - G_{c}^{'t} $.
The first set, $ G_{c}^{t}$, is constructed such that 
$ \forall n \in G_{c}^{t}, n \notin S_{C \setminus{c} }^{t} $.
That is, $ G_{c}^{t}$ contains nodes participating only in blockchain~$c$ at $t$.
Conversely, set $ G_{c}^{'t} $ contains the \textit{overlapping nodes}, \ie~those participating in blockchain~$c$ \textit{and at least another blockchain} $ c' \in C \setminus{c}$, at the same time $t$.

\subsection{Overlapping Network Entities}

A first approach in finding overlaps between BCs is by looking into our aggregated data-set, $\textbf{S}$ and count how many nodes and edges (\ie~pairs of endpoints), appear in more than one BC network, regardless of time.
Table~\ref{tab:overlaps-all} shows the summary of these results.
Evidently, there exists a significant number of network entities (both nodes and edges) that reside in more than one BC 
network.

\begin{table}[t]
\caption{Edge and Node overlaps (aggregated). 
$ON$: number of networks where a unique entity (node or edge) was found to be overlapping, regardless of time }
\label{tab:overlaps-all}
\begin{tabular}{|r|r|r|r|r|}
\hline
                & $ON$=2& $ON$=3& $ON$=4& $ON$>=5   \\
\hline
\textbf{Nodes}  & 34814 & 3909  & 1489  & 779       \\
\textbf{Edges}  & 143577& 11034 & 1958  & 222       \\
\hline
\end{tabular}
\vspace*{-4mm}
\end{table}

A second step is to investigate whether overlapping entities occur frequently or sporadically across time.
For this, we focus on the nodes and count all overlapping nodes in each $S_{c}^{t}$.
In Figure~\ref{fig:overlaps-per-ts}, for each BC network $c$, we plot the ratio 
of $|G_{c}^{'t}|$ over $|S_{c}^{t}|$, \ie~the number of overlapping nodes in snapshot $c$ 
over the total number of nodes in the snapshot.
We observe that in all BC networks, there is almost a constant percentage of overlapping nodes that join more than one BC
network. Therefore, from this and the previous results, we can state that overlaps exist between BC networks, they are significant and occur systematically through time.

\subsection{Structural Properties}
In this paragraph, we study the properties of overlapping nodes, compared to the rest of the network.
The main idea is to check for any differences between sets $G_{c}^{t}$ and $G_{c}^{'t}$ that can be supported with statistical significance. For the following analysis, we focused on the graph metrics presented in Sec.~\ref{sec:results-net}. 
Specifically, and for each BC network, we compared distributions of in-degree, out-degree, betweenness centrality,  clustering coefficient, and page-rank. Since we already found that some of these metrics (degree distributions) are  highly skewed (see scale-free property fitting paragraph), we perform a normality test on all metrics to decide on the  statistical method to be used for the comparison.
The normality test confirmed that the distributions of all metrics are in fact not normal. Normality was checked by performing the Shapiro-Wilk test, provided by the SciPy package~\cite{scipy}. Since we wish to compare distributions of non-normal data, a non-parametric test is needed.  Using the same package, we performed the 2-sample Kolmogorov-Smirnov test (KS-test) between all pairs of $G_{c}^{t}$ and $G'_{c^{t}}$, for all $c \in C$ and all $t$.

The KS-test returns the test statistic, $D$, which is the maximum distance between the CDFs of the two samples. It also returns the \textit{p-value} for the hypothesis test. If the test statistic $D$ is small, or the \textit{p-value} is higher than the selected statistical level $\alpha$ (\eg~0.05), then we cannot reject the null hypothesis that the distributions of the two samples are the same.
Our results indicate that in all networks, there is a significant distance between the metrics' distributions among groups $G_{c}^{t}$ and $G_{c}^{'t}$. Interested to see if there exists a metric that describes this difference better than the others, we looked into our results for the metric that gives significant $D$ values consistently.
 \begin{figure}[t]
\small
  \centering
  \subfloat[\centering Overlaps]{
    {\includegraphics[width=0.5\linewidth]{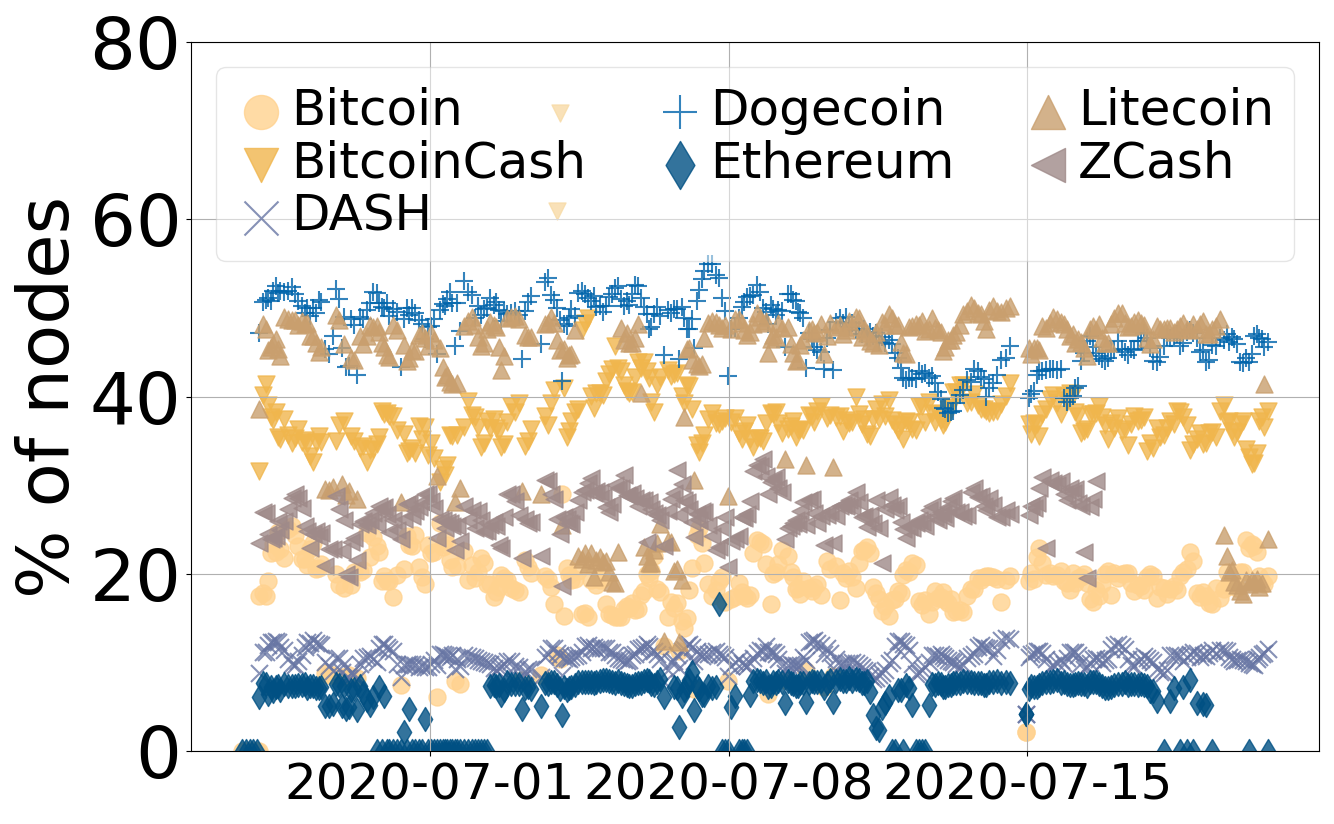}}
    \label{fig:overlaps-per-ts}}%
  \subfloat[\centering KS-Test results]{
    {\includegraphics[width=0.5\linewidth]{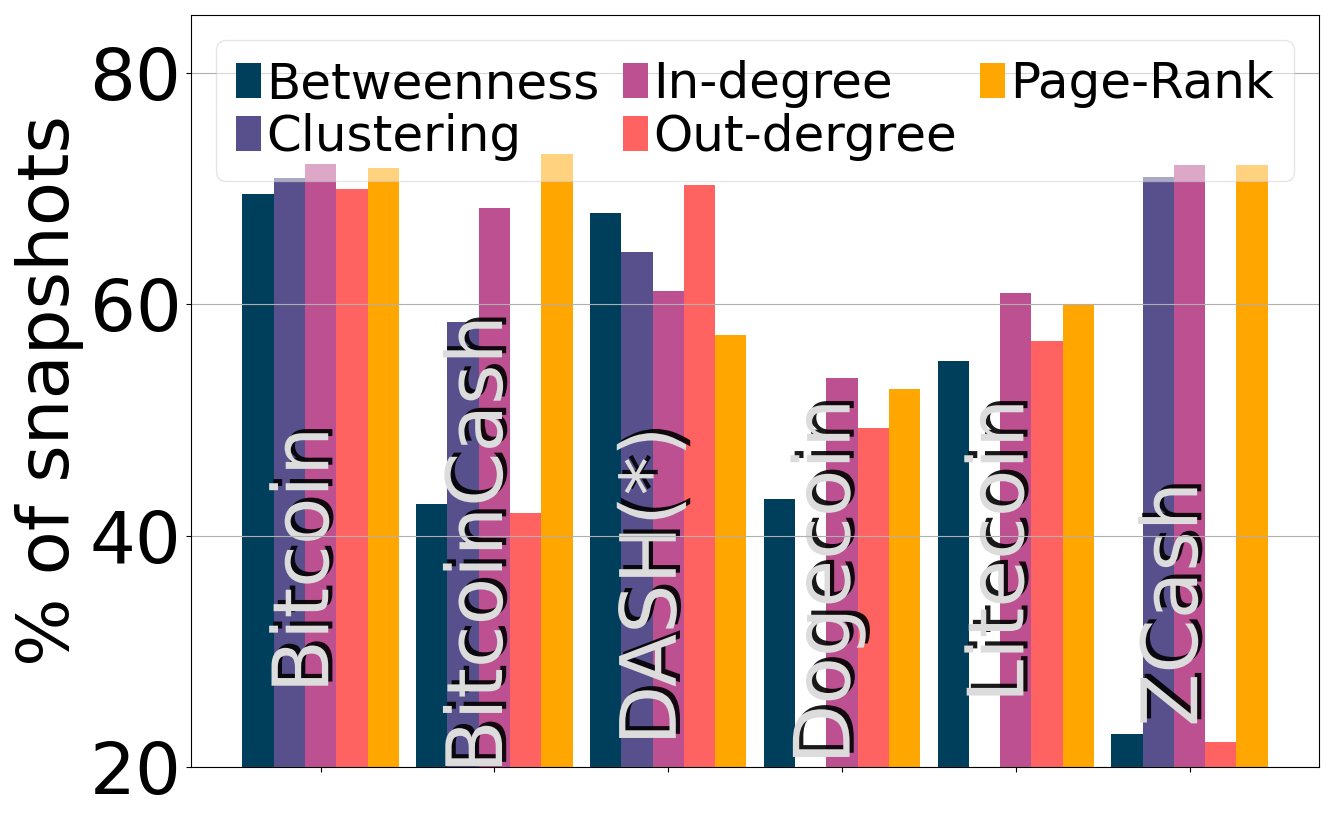}}
    \label{fig:ksresults}
  }%
\caption{
a) Percentage of nodes that were found in more than one BC network at the same timestamp.
X-axis indicates the timestamp.
b) Percentage of snapshots where the KS-test indicates a significant distance between the distributions of overlapping and non-overlapping nodes.}
\label{fig:overlaps}
\vspace*{-4mm}
\end{figure}
In Figure~\ref{fig:ksresults}, we plot the percentage of all snapshots of a given BC network $S_{c}^{t}$ where the p-value of the test is small enough, \ie~\textit{p-value}$< 0.05$, and $D > 0.1$.
The plot indicates that the distributions between the two groups are often non-equal.
We can also observe that in most networks the in-degree and page-rank metrics are the ones found more frequently, meaning that these metrics capture the differences between $G_{c}^{t}$ and $G_{c}^{'t}$, more often than the other metrics.

We also found that the CDFs of overlapping nodes are lower than non-overlapping nodes, meaning that the metrics of 
overlapping nodes are statistically higher. An exception in this finding was the Dash network (indicated with an * in 
Fig.~\ref{fig:ksresults}), where the opposite is true: in Dash, overlapping nodes have lower metrics than the rest.
A key takeaway from this test is that overlapping nodes are in fact different from other nodes. 
Although this test cannot serve as a proper classifier, it answers our question that overlapping nodes have in fact different properties from the rest of the nodes in a BC network.

\vspace*{-2mm}
\section{BC Longitudinal Evolution} \label{sec:results-temp}
In this section we address \textit{RQ3}, which concerns the temporal behavior of the BC networks and how they change over time. 
Permissionless BC networks are open, and nodes may freely come and go.
This independent arrival and departure of nodes creates the collective effect known as churn. Churn in Bitcoin has been studied by Imtiaz \etal~\cite{btc_churn}. A recent study by Kiffer \etal, explores Ethereum's overlay network in detail, including churn~\cite{kiffer-2021-ethereum}. Stutzbach and Rejaie made an in-depth study of churn in P2P networks~\cite{churnp2p}. 
In their work, they recognize that one of the most basic properties of churn is the \textit{session length} distribution, which describes how long each node participates in the system each time it connects. 
A \textit{session} begins when a node joins the network and ends when the node disconnects. 
In this section, we analyze the node dynamics observed in our collected datasets, by looking into the session length distributions.

\mysubsubsection{Limitations.}
As mentioned in Section~\ref{sec:methodology-details}, the snapshots were collected two hours apart. All peers that appear in a snapshot are considered to be online for the whole period.
This poses a limit to our longitudinal granularity; we cannot capture short-lived connections, i.e., we do not know whether a node came up multiple times within a two-hour period. Since disconnections during a snapshot period are missed, two sessions may look like one longer session. Moreover, very short sessions might not be observed at all, contributing to a bias towards longer-lived connections.

Additionally, due to maintenance operations, snapshots 95 and 253 were not purged to disk in a timely manner~(see Section~\ref{sec:methodology-details}). This resulted in two snapshots with duration of six and nine hours, respectively (timestamps 2020-07-03T07:03:22 and 2020-07-14T14:12:12).
We decided not to include these snapshots in the analysis. Furthermore, to better capture evolution dynamics, we split our dataset in 4 periods, excluding those two snapshots. Each of the four periods has an approximate duration of 5 and a half days:

\begin{itemize}
    \item \textit{period-0 (p0):} data from 2020-06-27 to 2020-07-02.
    \item \textit{period-1 (p1):} data from 2020-07-03 to 2020-07-08.
    \item \textit{period-2 (p2):} data from 2020-07-09 to 2020-07-14.
    \item \textit{period-3 (p3):} data from 2020-07-14 to 2020-07-20.
\end{itemize}

In our longitudinal analysis we decided to use the subset of \textit{reachable peers} of each snapshot, \ie peers that respond to our requests. This is preferred, since we cannot distinguish between peers that are offline and peers that are alive but behind NATs or firewalls.
Finally, during the last days of our measurement period, Zcash's network protocol was updated to a new version, incompatible with the previous one.
In effect, we could not capture any node in the Zcash overlay after the 16th of July 2020~\cite{zcashhheartwood}.

\subsection{BC Network Session Length}
\begin{figure}[t]
    \centering
    \subfloat[\centering]{\includegraphics[width=0.5\linewidth]{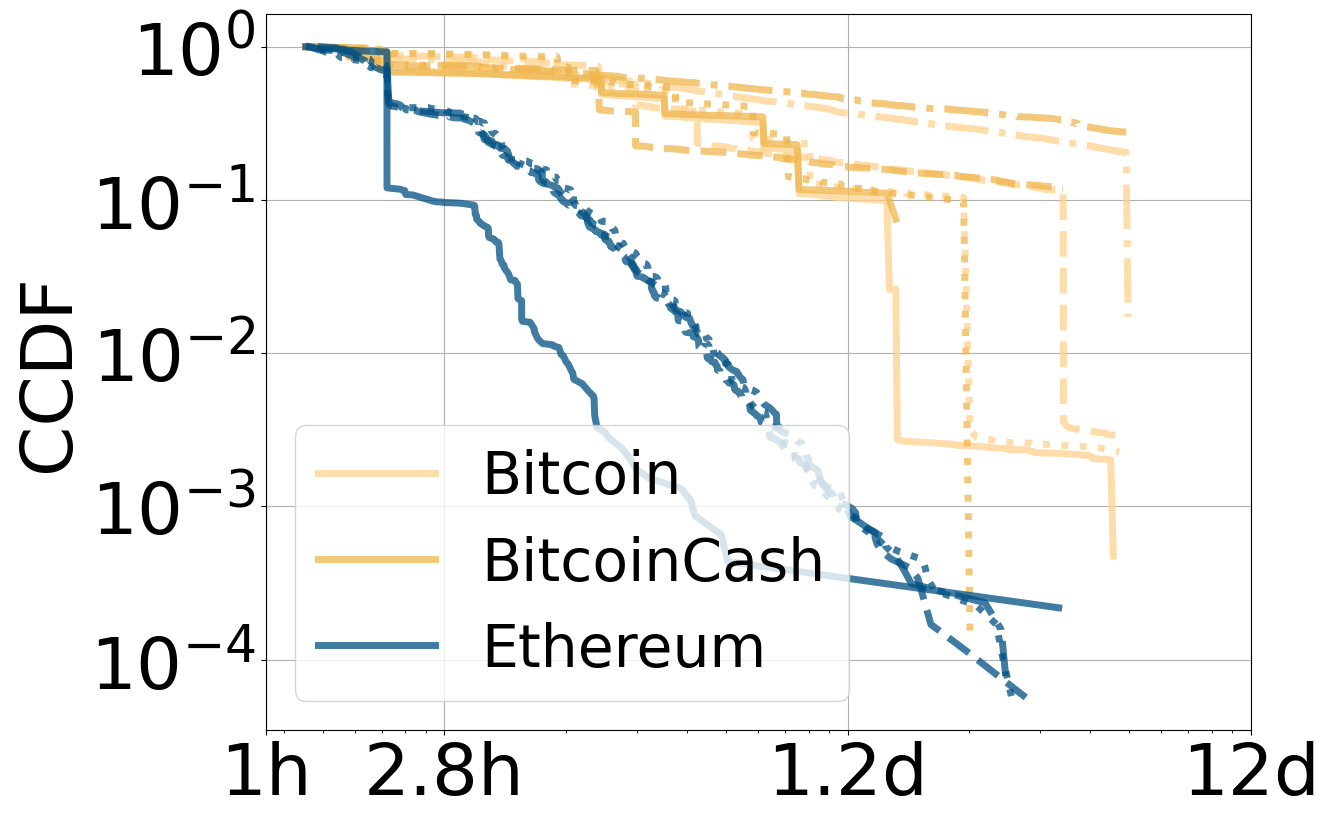}}
    \subfloat[\centering]{\includegraphics[width=0.5\linewidth]{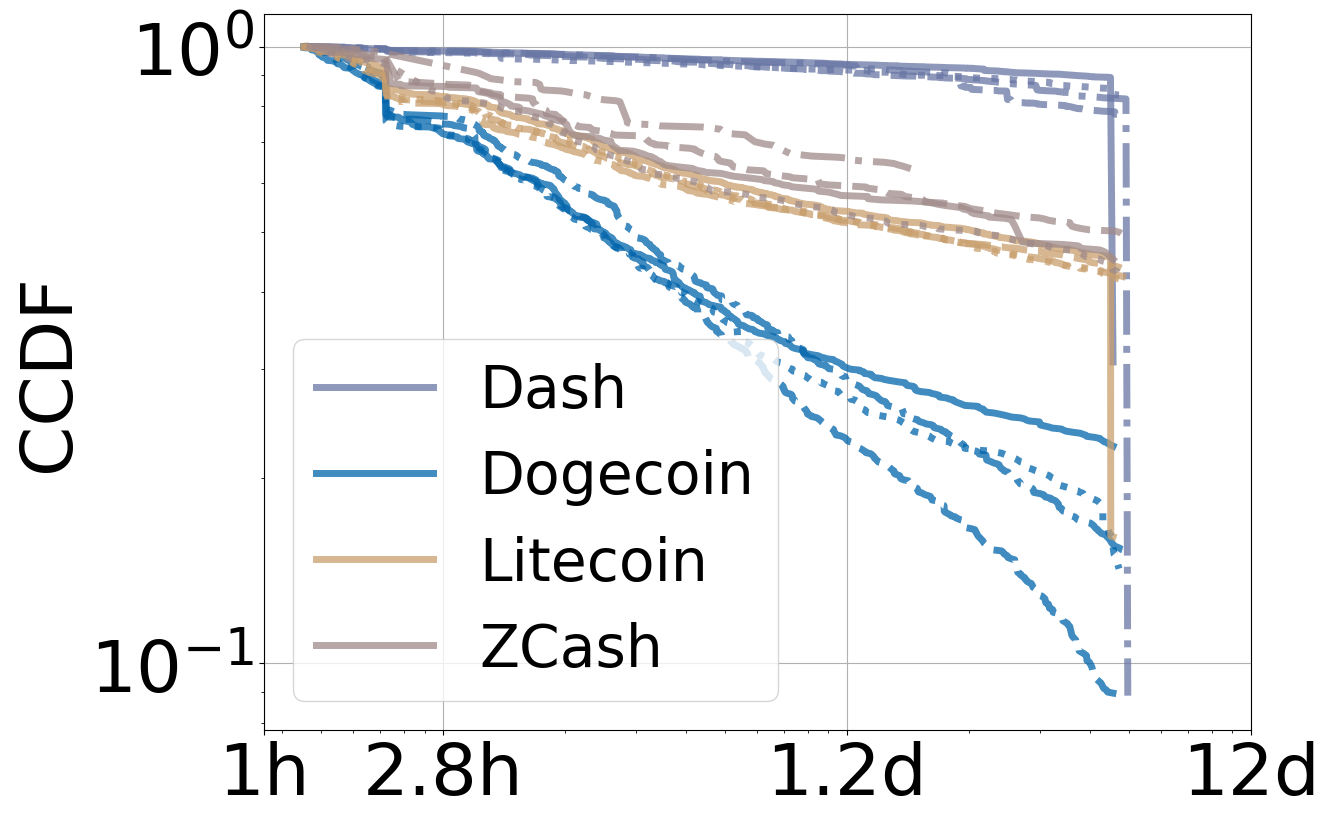}}
    \caption{CCDF of session lengths per BC network studied. Solid lines: 1st period; Dotted: 2nd; Dashed: 3rd; Dashdotted: 4th.}
    \label{fig:longi-session-lengths}
    \vspace*{-3mm}
\end{figure}

In Figure~\ref{fig:longi-session-lengths}, we plot the CCDF of session length across the four periods defined above, for each overlay.
As can be seen, the distributions for Dash, Dogecoin, Litecoin, and Zcash are very similar for all periods.
This suggests that in these overlays, the distribution of session lengths does not change significantly over time. Additionally, sessions seem to be consistent across these overlays.
In Bitcoin and BitcoinCash, we also observe that session lengths change over time.
Nevertheless, across the two systems, the distributions are consistent, with Bitcoin having a small percentage of nodes with sessions longer than BitcoinCash.
In Ethereum, we observe that most sessions are short.
In fact, fewer than half sessions are less than six hours.
Also, there is an obvious shift in session length distribution between the first and second period, which remains unchanged during the third period.
Finally, Dash has the highest percentage of long sessions, with 80\% of sessions lasting more than 4 days.

Prior studies\cite{btc_churn} suggest that session lengths exhibit a behavior similar to a heavy-tailed distribution. To test how well the session length distribution of each overlay can be modeled, we calculate the best fit using the same procedure as in Section~\ref{sub:scalefree}, while studying the scale-free property. In Table~\ref{tab:session-len-pl-fit}, we report the best fit for each period, for all BC networks.
Dash is best modeled by a stretched exponential in all periods, while Dogecoin and Litecoin are best fitted by a power-law with exponential cutoff.
The rest of the networks are mostly fitted by a power-law with exponential cutoff.

In summary, the studied BC networks vary in their temporal characteristics. The largest ones, namely Bitcoin, BitcoinCash, and Ethereum, have observable differences in session length distributions between different periods. In these networks, most sessions last less than a day and some sessions are week-long. The observed networks are better described by power-law distributions with exponential cutoff.

\begin{table}[t]
  \centering
  \caption{Session length best-fit.
    $PL$: power-law;
    $PLEC$: power-law with exponential cutoff;
    $SE$: stretched exponential.
    }
    \begin{tabular}{|l|l|l|l|l|}
    \hline
     &
      period-0 &
      period-1 &
      period-2 &
      period-3 
      \\
      \hline
    Bitcoin &
      SE   &
      PLEC &
      PLEC &
      PLEC
      \\
      \hline
    Bitcoin Cash &
      SE   &
      PLEC &
      PLEC &
      PLEC
      \\
      \hline
    Dash &
      SE &
      SE &
      SE &
      SE
      \\
      \hline
    Dogecoin &
      PLEC &
      PLEC &
      PLEC &
      PLEC
      \\
      \hline
    Ethereum &
      PLEC &
      PL &
      PL &
      PLEC
      \\
      \hline
    Litecoin &
      PLEC &
      PLEC &
      PLEC &
      PLEC
      \\
      \hline
    Zcash  &
      SE   &
      PLEC &
      SE   &
      SE
      \\
      \hline
    \end{tabular}%
  \label{tab:session-len-pl-fit}%
  \vspace*{-3mm}
\end{table}%

\subsection{Degree Vs. Session-length}
\label{sec:degree-seslen-corr}
To determine whether the node session-length correlates with a node's degree, we calculated the Spearman rank-order correlation~\cite{Kokoska-Spearman}, using the SciPy~\cite{scipy} package.

\begin{figure}[t]
    \centering
    \subfloat[\centering]{\includegraphics[width=0.5\linewidth]{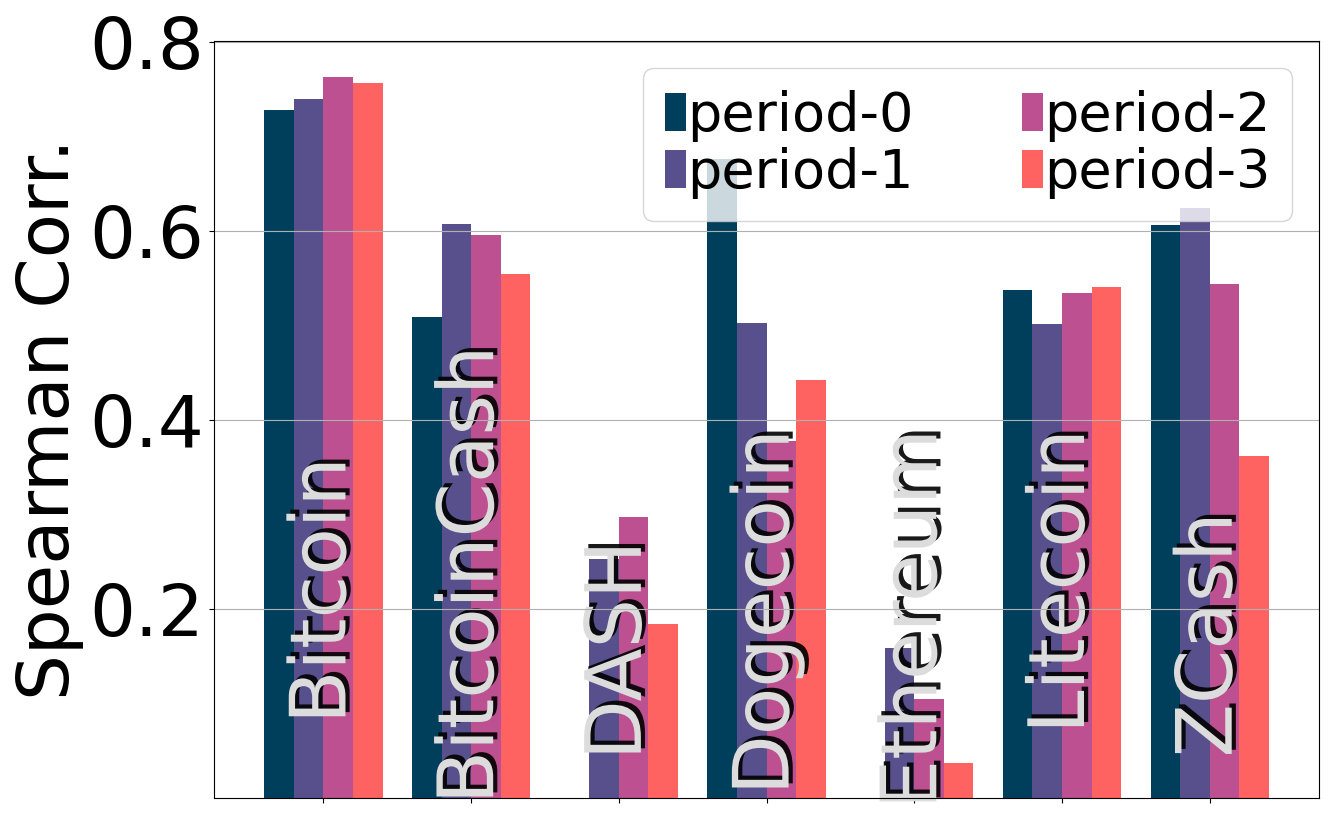}
    \label{fig:seslen-degree-spearman}}
    \subfloat[\centering]{\includegraphics[width=0.5\linewidth]{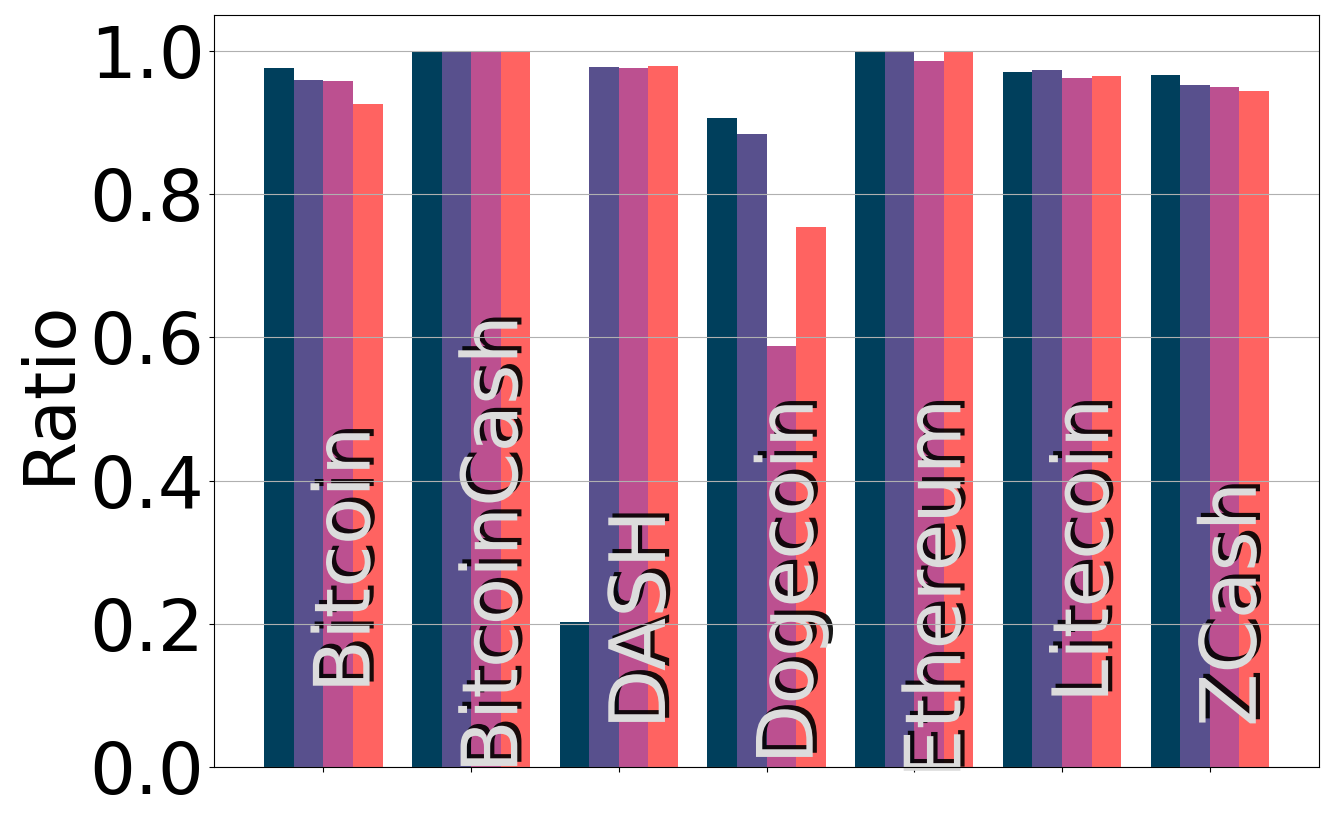}
    \label{fig:seslen-degree-ratio}}
    \caption{(a) Spearman Correlation between session-length and node-degree. (b) Percentage of high up-time nodes that are also high degree nodes.}
    \vspace{-4mm}
\end{figure}

In Figure~\ref{fig:seslen-degree-spearman} we plot the calculated correlation coefficients between node session-length and node degree for each overlay, per period. 
All correlation coefficients are statistically significant with p-values lower than 0.02, except for the first periods of Dash and Ethereum, which are not plotted.
The coefficients vary between the four periods and their values point towards moderate to strong correlation in all overlays. In Ethereum and Dash we observe the lowest correlations.
We further study the correlation in Ethereum by breaking each period into \textit{low} and \textit{high} degree nodes groups and computing the correlation coefficients per group.
The resulting correlation coefficients for higher degree nodes are again positive and higher, between 0.26 and 0.52.
The coefficients for the set of lower-degree nodes are negative, indicating a non-linear correlation between up-time and node-degree in Ethereum.
Similar observations are made in Dash as well, with high degree nodes having strong positive correlation and low-degree nodes with moderate negative correlation.
The relation between up-time and node-degree is further studied in Section~\ref{ssec:results-uptime-robust}

\section{Network Resilience to Attacks} \label{sec:results-robust}
In this section, we try to answer our final research question, \textit{RQ4}. 
How resilient are the BC networks to targeted attacks? How do shared (overlapping) nodes affect network resilience to such attacks? Finally, what happens when high up-time nodes are targeted by an attacker? To start this investigation, we first describe the attack model. Then, we define four strategies that an attacker could employ to partition a BC network, and we evaluate the efficacy of each strategy. 

\subsection{Attack Model}
An adversary may have various motives to attack a BC system. In this work, we specifically study attacks on the underlying topology of BC networks with the goal to impair the network's main functions. Specifically, we define the following two goals of the attacker.
\begin{enumerate}
    \item Split the BC network into two or more partitions
    so that no information flows among them. 
    \item Disturb the information propagation mechanisms by introducing intolerable delays.
    Such delays can typically increase the time to reach consensus among all participants and create a split in the application layer of a BC system.   In fact, propagation delays are known to be key contributors towards BC forks~\cite{Decker2013}.
\end{enumerate}

Such attacks would impair a BC network's main functions, potentially breaking user trust in the system. Attackers with no stake in the particular Blockchain's currency could be highly motivated to perform such attacks. To measure the efficacy of each attacker goal, we use three metrics: a) the size of the largest weakly connected component, b) the number of connected components, and c) the network diameter. To this end, we consider the following attack strategies:
\begin{enumerate}
    \item Targeted attacks on unique nodes, based on a selected network metric. We test out-degree and betweenness centrality, but others can be employed.
    \item Targeted attacks on nodes overlapping across more than one BC network. Nodes are ranked on their betweenness centrality.
    \item Random attacks using random node removals emulate failures that can occur in the network in random fashion and are used as a baseline.
    \item Attack minimum-cut edges, in order to partition the network by removing edges that are positioned in key places in the graph.
\end{enumerate}

\begin{figure*}[ht]
  \centering
  \subfloat
  {{\includegraphics[width=0.33\linewidth]{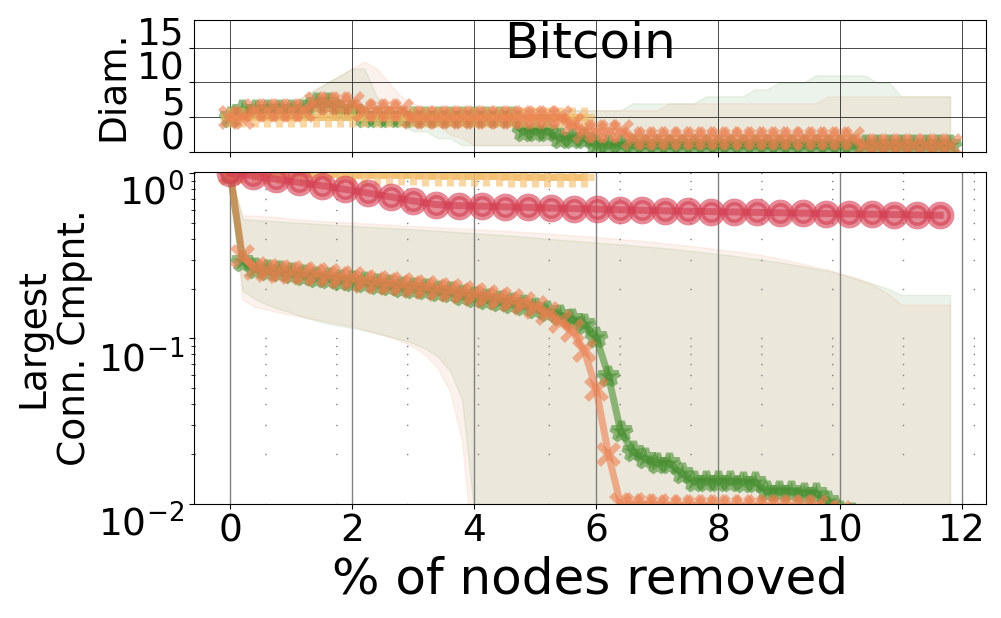}}}%
  \subfloat
  {{\includegraphics[width=0.33\linewidth]{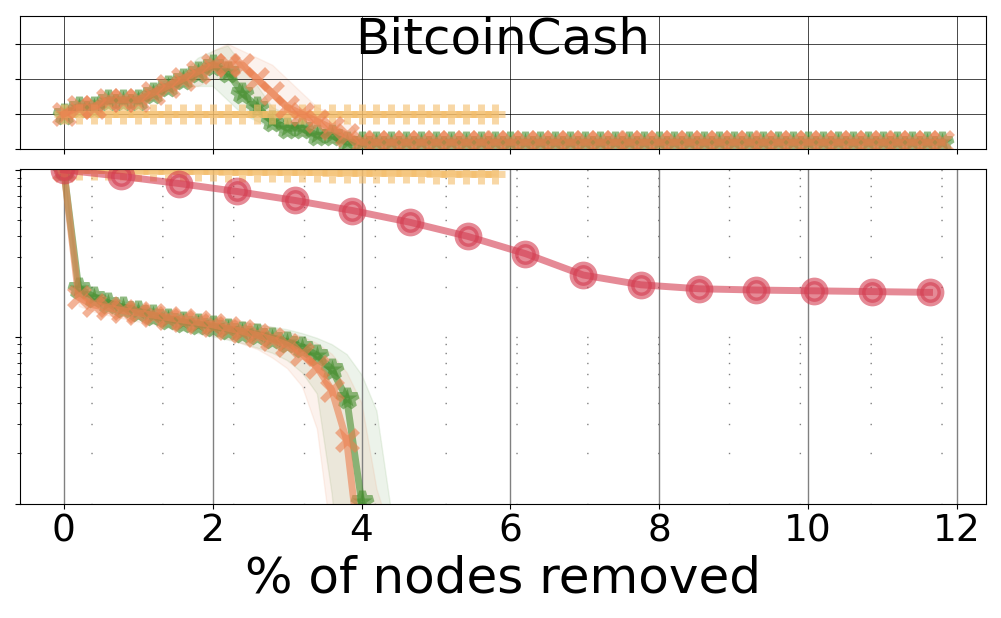}}}%
  \subfloat
  {{\includegraphics[width=0.33\linewidth]{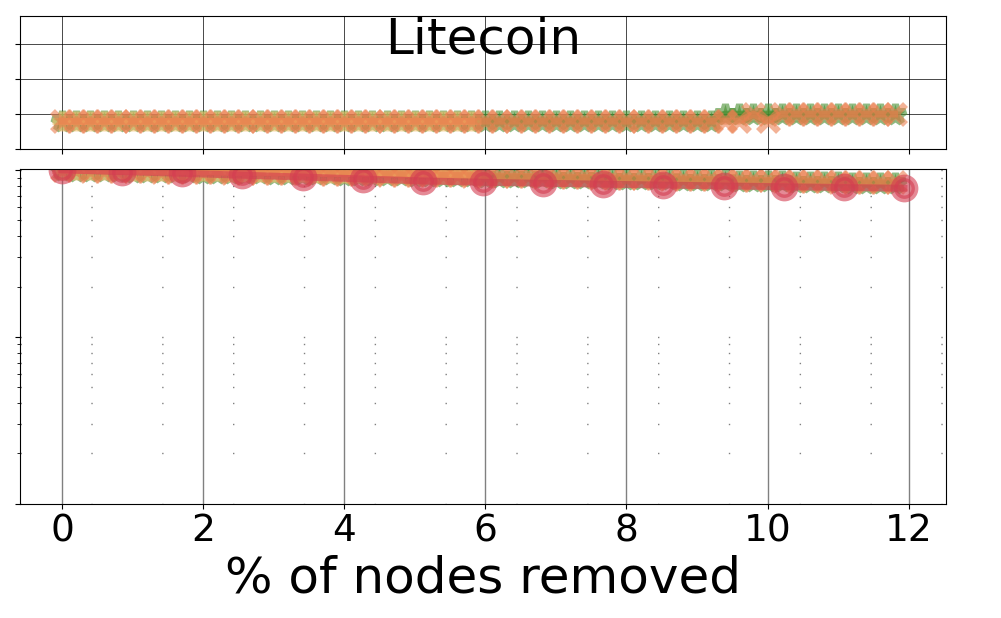}}}%
  \:
  \subfloat
  {{\includegraphics[width=0.33\linewidth]{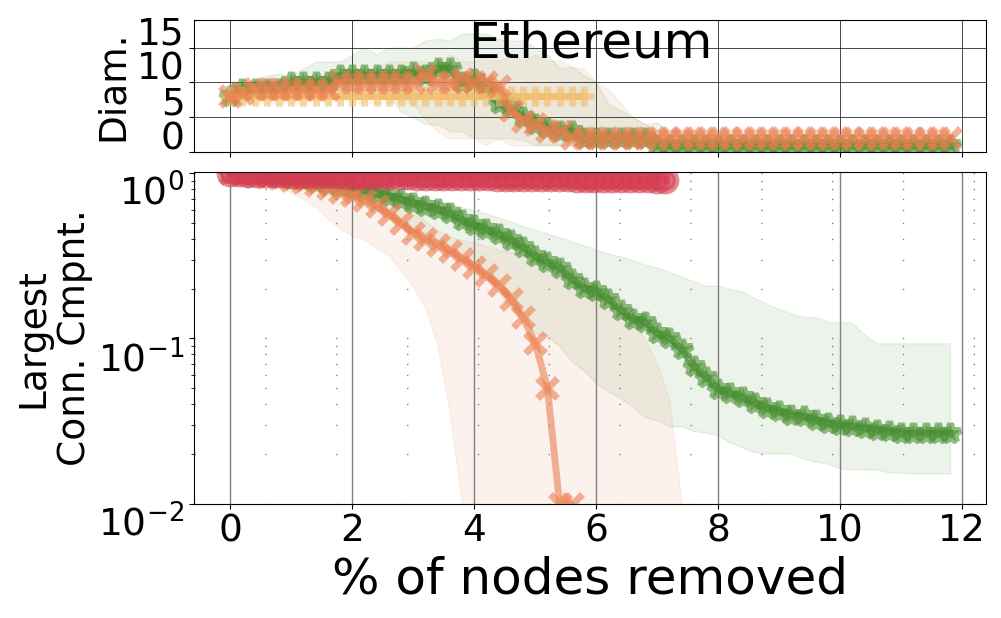}}}%
    \subfloat
  {{\includegraphics[width=0.33\linewidth]{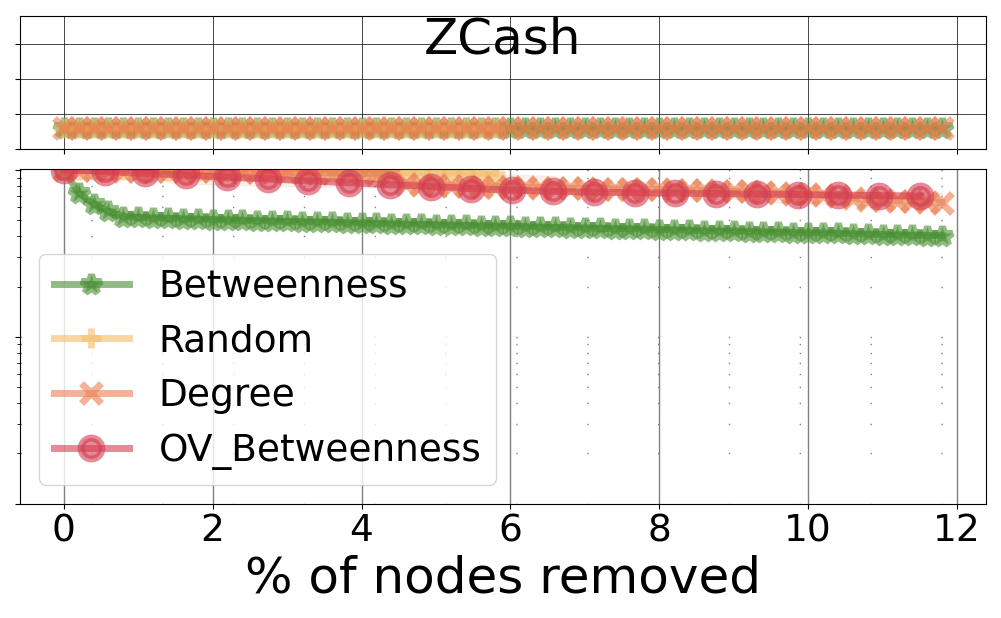}}}%
    \subfloat
  {{\includegraphics[width=0.33\linewidth]{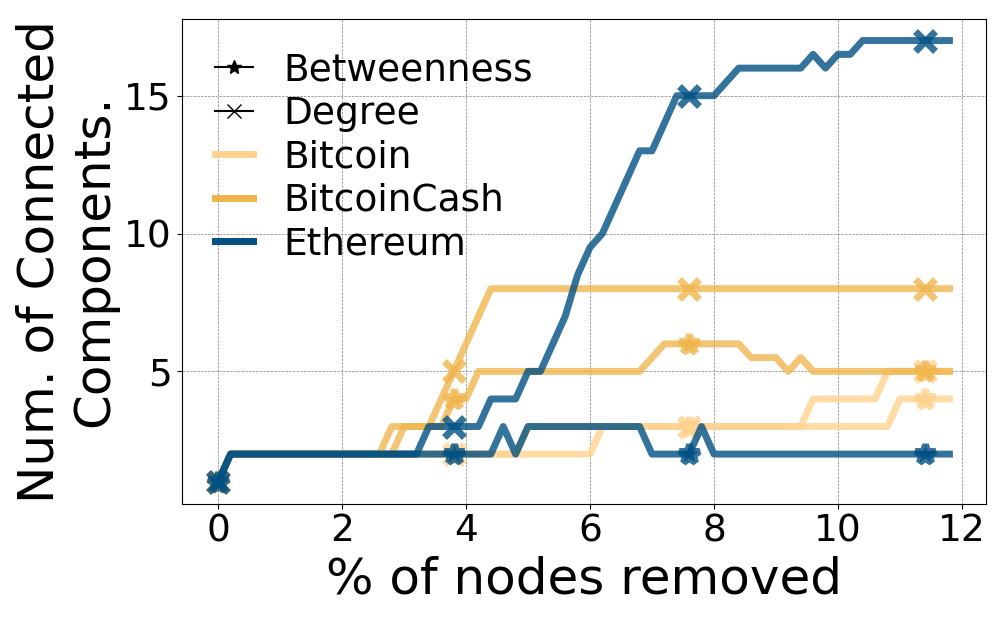}}}%
    \caption{Evolution of the approximate diameter (upper part) and size of largest weakly connected component (lower part) when the network in under targeted attack.
    The X-axis reports percentage of nodes removed.
    The lines correspond to the median value across all snapshots.
    The shaded area indicates values between 1st-3rd quartile.
    Orange x: Out-degree of unique nodes;
    Beige +: Random unique nodes;
    Green *: Betweenness of unique nodes;
    Red o: Betweenness of overlapping nodes;
    Dash and Dogecoin exhibit a similar shape with Litecoin and are not included due to space limitations.
    The rightmost-bottom graph plots the evolution of number of connected components during the same experiment.
    }
\label{fig:percolation-high}%
\end{figure*}

\subsection{Targeted Node Attacks}
\subsubsection{Individual BC Networks.}
The removal of a node simultaneously cuts all its adjacent links, hence it is more efficient for the attacker compared to targeted link removals. We focus on how to remove nodes in the most efficient way to minimize the amount of node removals necessary to cause a disruption. A node can be removed from the network through various means, including DoS attacks. We follow a static procedure, in the sense that each node is given a static priority of removal, based on a chosen metric. For instance, when using the out-degree metric, the higher the degree, the higher the importance of the node to be attacked. After a node is removed, priorities are not recalculated. We remove nodes from the network one by one, following the given priority. After each removal, we calculate the size of the largest weakly connected component and the approximate diameter of the resulting graph. We report the effectiveness of the two node ranking metrics (betweenness centrality and out-degree), and compare with the baseline random removal strategy. We perform the procedure described for all snapshots per BC  network. Due to the high number of graphs collected, we stop execution after removing 12\% of nodes per snapshot.

As can be seen in Figure~\ref{fig:percolation-high}, 
in Bitcoin and Bitcoin Cash, the betweenness and out-degree metrics have roughly the same shape. The size of the largest connected component falls significantly after the removal of just a few nodes. Further removal of nodes shrinks the size gradually until a threshold where the connected component falls abruptly to 1\% of its initial size. This occurs upon removal of 6\% and 4\% of the nodes respectively. 
Such a behavior is also found in the Internet~\cite{magoni2003tearing}.
This finding may not seem very worrisome, since the reported percentages correspond to a few thousand nodes. Nevertheless, closer inspection (not shown in Figure) of these two networks, indicates that the removal of the first 5 nodes, reduces the size of the largest connected component by 60\%, which is alarming. In contrast to the size of the largest component, the network diameter starts increasing earlier in this process
and this effect is more pronounced in Bitcoin Cash. 

Dash, Dogecoin and Litecoin seem equally resilient to random and targeted attacks.
The size of the largest component drops linearly with the number of nodes removed and their diameter is not significantly affected. We can attribute their resilience to their structural characteristics discussed earlier. All three networks have a very large strongly connected component, high clustering, and high average degree (Table~\ref{tab:all-metrics}).

In Ethereum, the out-degree strategy is more potent. In contrast to Bitcoin and Bitcoin Cash, the size of the largest component does not drop initially. After removing 2\% of the nodes, the size drops gradually up to a threshold, close to 5\%, where its size falls abruptly to 1\%. The network diameter starts increasing early in the process, but not as quickly as in Bitcoin Cash. 

When targeting high betweenness nodes in Zcash, the largest component initially falls abruptly. Similar to Bitcoin, the first removal of nodes reduces the largest component by 40\%. When 4\% of the nodes are removed, the largest component drops to 50\% of its initial size, and then shrinks almost linearly. Targeting high out-degree nodes is less damaging in Zcash. More than 5\% of the nodes need to be removed in order to observe a 20\% reduction of the largest component. We summarize our findings in Table~\ref{tab:resilience-summary}.

\begin{table}[h]
  \centering
  \caption{Resilience of BC networks to targeted node attacks. We report the number and percentage of nodes that, when removed reduce the largest component to 50\% and 1\% of its initial size respectively.}
    \begin{tabular}{|p{6.3em}|c|c|c|c|c|c|c|}
\hline
  \multicolumn{1}{|l|}{Network} &
      Bitcoin &
      \multicolumn{1}{|p{5.4em}|}{Bitcoin Cash} &
      Dash &
      Dogecoin &
      Ethereum &
      Litecoin &
      Zcash
\\
\hline    
  \# of Nodes (50\% reduction) &
      5 &
      4 &
      - &
      - &
      300 &
      - &
      25
\\
\hline    
  \% of Nodes (99\% reduction) &
      6.5\% &
      4\% &
      >12\% &
      >12\% &
      5.5\% &
      >12\% &
      >12\%
\\
\hline   
  \end{tabular}%
  \label{tab:resilience-summary}%
  \vspace{-3mm}
\end{table}%

\subsubsection{Overlapping BC Networks.}
The last part of our research question \textit{RQ3} asks how overlapping nodes can affect resilience of BC networks to targeted attacks. In order to answer this, we repeat the experiment of the previous paragraph with a small variation.
From each set $S_{c \in C}^{t}$, we remove all $G_{c}^{t}$ sets. 
This new set, $S_{c \in C}^{'t}$, contains all nodes that participate in more than one 
networks at the same time $t$.
We then order the unique elements of $S_{c \in C}^{'t}$, in descending order of their maximum normalized betweenness centrality. Since a node participates in more than one BC networks, we sort nodes based on the maximum value they have across all snapshots at $t$.
We perform normalization using the Min-max method, per snapshot.
We then proceed by removing the nodes in $S_{c \in C}^{'t}$ according to their rank, from each snapshot $S_{c}^{t}$ at the same time $t$. Nodes are removed in the same order from all snapshots.

The results of targeting overlapping nodes first, are plotted in  Figure~\ref{fig:percolation-high} using red circles. The plot reports the average change in the largest connected component over all snapshots $S_{c}^{t}$. Clearly, this strategy is less effective compared to the strategies used earlier, which target top central nodes within a specific BC network. Nevertheless, it provides an adversary the benefit of attacking multiple BC networks simultaneously. One interesting finding is that Litecoin is more susceptible to this kind of attack compared to attacks focusing on single BC node metrics. This is partly explained by the fact that Litecoin has one of the highest percentages of overlapping nodes (see Fig.~\ref{fig:overlaps-per-ts}).
Closer inspection of the data at hand shows that an attacker is able to shrink the largest connected component of Bitcoin Cash, Bitcoin, Zcash, and Litecoin networks by 70\%, 40\%, 25\% and 20\% respectively. This demonstrates that, by targeting overlapping nodes, a powerful adversary can still mount a successful partitioning attack in \emph{4 different networks at the same time}.


\subsubsection{High up-time nodes.}
\label{ssec:results-uptime-robust}
As already mentioned in Section~\ref{sec:degree-seslen-corr},
there is positive correlation between a node's session-length and its degree.
Further inspection of the data revealed that longer-lived nodes are almost always high-degree nodes.
This is explained by the fact that the longer a node is up, the more peers it is able to discover.
In Figure~\ref{fig:seslen-degree-ratio}, we plot the ratio of high session-length nodes that are also high-degree nodes.  For each snapshot, we extract the top 10\% nodes ranked by their out-degree.
We then set the least up-time of these nodes as a threshold.
By selecting the nodes with an up-time longer or equal to that threshold, we calculate the Jaccard similarity between the two sets of nodes, namely \textit{Long-session nodes} and \textit{High-degree nodes}.
This is directly related to our targeted node removal experiment. That is, if we were to repeat the experiment, using a strategy based on nodes' up-time, we would end-up selecting a similar set of nodes with the \textit{highest-degree} strategy already studied.
This leads us to expect similar results in the networks' resilience.
The observed relation between a node's up-time and degree could be used to locate nodes of interest in settings that employ topology hiding features, as is the case with the latest Bitcoin reference client (see Section~\ref{sec:bitcoin-topology-hiding}).

\subsection{Minimum Edge Cuts}
Targeting minimum cut edges does not have a significant effect in the networks’ state and requires the removal (or disruption) of a considerable number of network links. In Appendix~\ref{app:edge-cuts} we discuss our approach and list our results when employing this strategy.

\section{Summary - Key Takeaways}
\label{sec:results-summary}

We have investigated the network characteristics of the BC overlays. Furthermore, we studied the presence of overlapping nodes and their properties. Lastly, we investigated several targeted attacks on the connectivity of BC networks. Next, we summarize the important findings in our analysis:
\begin{itemize}
\item  BC overlays vary significantly in their densities, with larger networks being less dense.
\item Their dynamics are manifested by significant variations found in the degree distribution per snapshot.
\item Their clustering coefficient distributions are similar to other real networks, and differ from random networks with similar size and average degree. 
\item They are well-connected, and their degree distributions belong to the exponential family. Despite their low diameters and small average shortest path length we did not find evidence that these networks satisfy the small-world property.
\item At all times, a non-negligible amount of overlapping nodes exists between various BC networks.
\item We have strong statistical evidence that overlapping nodes differ in their properties and metrics' distributions from the rest of the nodes within a BC.
\item BC overlays have varying temporal characteristics and there exists a strong positive correlation between a node's session-length and its degree.
\item BC overlays are very resilient against random failures but targeted attacks can considerably affect their connectivity; some networks can be partitioned by the removal of less than 10 well-connected nodes.
\item A powerful adversary could disrupt at least 4 BC overlays by targeting overlapping nodes.
\end{itemize}
\vspace*{-3mm}
\section{Conclusion}
We provide a first and in-depth look into the overlay properties and structural resilience of seven prominent (in terms of capitalization) Blockchain (BC) networks. We use custom-made Blockchain crawlers to probe 32 million BC peers, obtain each peer's list of known peers, and extract their possible connections.  
We find that the structure of BC networks is unlike that of traditional networks (\eg~the Web), but shares some similarities. We also find that the graph properties vary substantially among the studied networks. 

Moreover, we discover that Bitcoin, Bitcoin Cash, Zcash, and Litecoin share a substantial amount of nodes. This fact hints at the efficacy of targeted attacks on a limited set of shared nodes  to disrupt more than one networks.  At the same time, 
we find that all networks are robust against random edge removal but some are vulnerable to targeted removal strategies. Importantly, the removal of very few nodes in Bitcoin, BitcoinCash, and Zcash, results in a significant reduction of the largest weakly connected component's size. 

Our results raise the alarm with respect to the resilience of the studied Blockchains against partitioning and message propagation delay attacks. We demonstrate that, by using our methodology, a deliberate and methodical attacker can uncover a small set of entities central to the topology and target them to substantially suppress message propagation in more than one BC network simultaneously. This highlights the need to employ measures to enhance network robustness (\eg~by increasing the number of connections between peers) rather than relying on topology hiding.

\bibliographystyle{ACM-Reference-Format}
\bibliography{main}

\appendix
\section{Ethics}
\label{app:ethics}
In this work we followed standard ethical guidelines~\cite{dittrich2012menlo,10.12688/f1000research.3-38.v2,DBLP:conf/imc/AllmanP07} for the collection and sharing of measurement data. We only collect and process publicly available data, make no attempt to deanonymize users or link people and/or organizations to their IP address. No personally identifiable information was collected.

While crawling the networks we only take part in the peer discovery mechanism of each network and gather IP addresses known to each node. Those addresses were only used to synthesize connectivity graphs on which our research was based. We did not try to identify any user by her IP address and no information was redistributed. In fact, our crawler created short lived connections to any discovered peer in the network and did not respond to any other requests except the expected initial handshake. We do not respond to any other messages or requests. In addition, we employed low bandwidth utilization to avoid resource exhaustion.
Our measurements did not cause any disruption or exposure of the BC networks under study.

Our results unveil particular nodes whose targeting has the potential to disrupt the overlay's operation. To prevent misuse of this portion of the results, we do not publish the IP address of any node in our dataset. We instead replace the IP address with a persistent random identifier and we privately maintain a private map of IPs to random identifiers for verification and reproducibility purposes.


\section{Methodology Assessment} \label{app:methodology-assessment}
We evaluate the efficacy of our method as follows.
We setup an unmodified Bitcoin reference node using the official Bitcoin core implementation\cite{releasenotesbtc}. After the initial deployment, we allowed the reference node to perform its initial bootstrap of the BC for one week. Subsequently, every ten minutes we retrieve the following information from the reference node: a) all inbound and outbound connections,
b) a copy of the \texttt{peers.dat} file, containing all known peers, and
c) the \texttt{ADDR} reply to a \texttt{GETADDR} probing message.
Looking into the collected data, we found:
\begin{enumerate}
\item After more than one week of operation, the \texttt{peers.dat} file of the reference client contained 41k IP addresses.
\item Some peers are more frequently included in \texttt{ADDR} replies than others.
By sending 20 \texttt{GETADDR} messages, we were able to retrieve 17k unique peers, instead of the expected 20k (1k per message), meaning that a group of nodes appears more frequently than others.
\item The number of outgoing connections of the reference client was between 8 and 12. Most of the outgoing connections were included in the node's replies to the crawler.
\end{enumerate}

Imtiaz \etal~\cite{btc_churn} studied churn in the Bitcoin network and showed that the majority of peers stays online for less than a day, and more than 95\% of nodes stay online for less than a week.
Thus, we counted the number of actual outgoing connections being included in an ADDR reply for four consecutive days. 
In all replies we are able to retrieve more than 80\% of the actual connections. 
Even after numerous restarts (during which the node's \texttt{ADDRMAN} gets enriched by asking DNS seeds and performing an initial peer discovery round) 
we retrieve more than 80\% of actual connections, on average.
We note that for this calculation, we filter out transient connections, \ie~connections that last for less than 2 hours.
From the above we conclude that our earlier proposed method can repeatedly retrieve a good representation of a node’s outbound connections.
The resulting graphs are synthesized by the \texttt{edge} sets collected. 
Although the precision is low, the collected graphs
contain the outbound connections of each node with high probability and may be considered as graphs of the overlay network topology.
The actual true connections of active nodes would then be spanning sub-graphs of the synthesized graphs, containing all vertices but less edges.

\section{Small-world Property}
\label{app:small-world}
The small-world phenomenon states if you choose any two individual nodes of a small-world graph, the distance between them will be relatively short, and definitely orders of magnitude smaller than the size of the network.
We examined all collected snapshots to see if they satisfy the small-world property, by calculating the $\omega$ metric 
proposed in~\cite{telesford2011smallworld}.
The metric is defined as $ \omega ={\frac {L_r}{L}}-{\frac{C}{C_l}}$
where $L$ and $C$ are the average shortest path and average clustering coefficient of the snapshot, respectively.
$L_r$ is the average shortest path for an equivalent random network and $C_l$ is the average clustering coefficient of an equivalent lattice network. The value of $\omega $ ranges between $-1$, when the network has lattice characteristics, to $+1$ when the network has random graph characteristics, with values near $0$ interpreted as evidence of small worldliness. 
We did not find evidence that the networks under study satisfy this property. Although we observe low average distances in all BC networks, they do not have high enough clustering coefficients to be considered as small-world. Indicatively, the $\omega$ values we calculated are greater than $0.5$ for Dash and Zcash. The rest of the networks have values greater than $0.8$.

\section{Minimum Edge Cuts}
\label{app:edge-cuts}
To compute the minimum edge cuts, we used the algebraic connectivity of the derived graphs. The algebraic connectivity of a graph is defined as the second smallest eigenvalue of its Laplacian matrix $L$, $\lambda_{2}(L)$ and is a lower bound on node/edge connectivity~\cite{fiedler73algebraic}. Since calculating the algebraic connectivity of a graph is computationally very expensive (\ie~>3 compute hours per snapshot), we analyse one snapshot per BC network. Using the computed eigenvector, we count how many edges are required to be removed to split the network in two parts, and compute their sizes and ratio of the two subnets (cut-ratio computed as number of nodes in the largest subnet over the total number of nodes). Results are presented in Table~\ref{tab:fiedler}.
\begin{table}[t]
  \centering
  \caption{Resilience of BC networks in edge and node removal. Critical threshold, \protect{$f_{c}$}, is the percentage of nodes that, when removed reduce the largest component to 1\% of its initial size.}
    \begin{tabular}{|r|c|c|c|c|}
    \hline
     &
      Edges &
      \multicolumn{1}{c|}{\multirow{2}[2]{*}{Cut Ratio}} &
      \multicolumn{1}{p{3.665em}|}{\#Nodes} &
      \multirow{2}[2]{*}{$f_c$}
      \\
     &
      Removed &
      &
      50\% split &
      
      \\
    \hline
    Bitcoin &
      5545(0.1\%) &
      0.815 &
      5 &
      6.5\%
      \\
    \hline
    Bitcoin Cash &
      10603(6.5\%) &
      0.511 &
      4 &
      4\%
      \\
    \hline
    Dash &
      1451(0.02\%) &
      0.995 &
      - &
      >12\%
      \\
    \hline
    Dogecoin &
      581(0.44\%) &
      0.990 &
      - &
      >12\%
      \\
    \hline
    Ethereum &
      2220(2.71\%) &
      0.972 &
      300 &
      5.5\%
      \\
    \hline
    Litecoin &
      544(0.08\%) &
      0.994 &
      - &
      >12\%
      \\
    \hline
    Zcash &
      363(0.33\%) &
      0.827 &
      25 &
      >12\%
      \\
    \hline
    \end{tabular}%
    \label{tab:fiedler}
\end{table}%
Most cuts are heavily unbalanced. Bitcoin Cash has an almost perfect cut, albeit by removing a high fraction of edges (6.5\% of edges or 10k edges out of total).
Bitcoin and Zcash are somewhat affected, by removing less than 0.5\% of their network edges. Overall, targeting minimum cut edges does not have a significant effect in the networks' state and would require the removal (or disruption) of a considerable number of edges connecting nodes.

\end{document}